\documentclass[preprint,12pt]{elsarticle}

\usepackage{amsmath,amsfonts,amssymb,amscd,amsthm,xspace}
\usepackage[scriptsize]{subfigure}
\usepackage{booktabs}
\usepackage{hyperref}

\journal{Nuclear Physics A}

\begin{document}

\begin{frontmatter}



\title{Lifetime of the $\eta^\prime$ meson at low temperature}


\author[rvt]{E.~Perotti}
\author[rvt,focal]{C.~Niblaeus}
\author[rvt]{S.~Leupold}
\address[rvt]{Department of Physics and Astronomy, Uppsala University, Box 516, 75120 Uppsala, Sweden }
\address[focal]{Oskar Klein Centre for Cosmoparticle Physics and Department of Physics, \\ Stockholm University, 10691 Stockholm, Sweden}

\begin{abstract}
This work constitutes one part of an investigation of the low-temperature changes of the properties of the $\eta'$ meson. In turn these properties are strongly tied to the $U(1)_A$ anomaly of Quantum Chromodynamics. The final aim is to explore the interplay of the chiral anomaly and in-medium effects. We determine the lifetime of an $\eta'$ meson being at rest in a strongly interacting medium as a function of the temperature. To have a formally well-defined low-energy limit we use in a first step Chiral Perturbation Theory for a large number of colors. We determine the pertinent scattering amplitudes in leading and next-to-leading order. In a second step we include resonances that appear in the same mass range as the $\eta'$ meson. The resonances are introduced such that the low-energy limit remains unchanged and that they saturate the corresponding low-energy constants. This requirement fixes all coupling constants. We find that the width of the $\eta'$ meson is significantly increased from about 200 keV in vacuum to about 10 MeV at a temperature of 120 MeV.
\end{abstract}


\end{frontmatter}


\section{Introduction}
\label{Intro}
In the limit where the masses of the three lightest quarks are set to zero (see \cite{Weinberg:1978kz,Gasser:1983yg,Gasser:1984gg,Scherer:2002tk} and references therein), the Lagrangian 
of the strong interaction possesses a chiral symmetry with group structure $U_L(3) \times U_R(3)$. The vector part of this 
symmetry gives rise to baryon-number conservation and the flavor multiplets of hadrons. 

The axial-vector part related to $U_A(3) = SU_A(3) \times U_A(1)$ is broken. The spontaneous breaking of $SU_A(3)$ leads to 8 
pseudoscalar Goldstone bosons. Finally, $U_A(1)$ is broken by the quantization procedure, the chiral anomaly. If it were not for 
the anomaly, $U_A(1)$ would be spontaneously broken, giving rise to a 9th Goldstone boson \cite{Kaiser:2000gs}. The anomaly 
provides a significant mass for the lowest-lying pseudoscalar flavor-singlet state. 
It also dictates the size of several low-energy quantities, for instance the photon-3-pion amplitude and the lifetime of 
the neutral pion, which would live much longer without the anomaly \cite{Wess:1971yu,Witten:1983tw,weinberg-textbook}. 

In reality, i.e.\ with non-vanishing quark masses, all these effects persist. Yet the then quasi-Goldstone bosons obtain finite 
masses and the pseudoscalar flavor-octet iso-singlet state mixes with the flavor-singlet state. The nine states emerging in 
this way are the three pions, the four kaons, the $\eta$ meson and the $\eta'$ meson. It is the latter, heaviest one 
that is closest tied to the chiral anomaly: it would be the ninth quasi-Goldstone boson generated by the spontaneous breaking of $U_A(3)$ if there was no anomaly.

Typically symmetries not realized at low temperatures become restored at some transition temperature. This is the case of the chiral symmetry of the strong interaction --- related to $SU_A(3)$ --- which is believed to be restored at a temperature $T_c\approx 170$ MeV (see \cite{Friman:2011zz} and 
references therein). At such high temperatures quarks and gluons are deconfined, giving rise to a new state of matter: the Quark-Gluon Plasma. This agrees with what we expect when we observe a phase transition: a change in symmetry. However what happens to the observables governed by the chiral anomaly in a strongly interacting many-body system, a ``medium'', has not yet been fully understood. Anomaly related amplitudes involving pions have been addressed in \cite{Pisarski:1996ne,Pisarski:1997bq}. 

It is then of great interest to investigate how the in-medium properties of the $\eta '$ meson change with temperature since they indirectly provide information about possible changes of the chiral anomaly in a thermal system ~\cite{Kapusta:1995ww,Kwon:2012vb,Sakai:2013nba}.  
One aspect, the low-temperature mass change, has been addressed for example in \cite{JalilianMarian:1997mb,Escribano:2000ju}. 
Yet, it might be somewhat oversimplifed to concentrate solely on possible mass changes. In-medium effects
are often much richer and more intriguing \cite{Kapusta:1993hq,Rapp:1999ej,Leupold:2009kz}. In particular at non-vanishing temperature, 
chiral restoration is accompanied by 
the transition from confinement to deconfinement. The decrease of the lifetime of a hadronic excitation due to 
multiple collisions can be interpreted as 
a precursor to deconfinement \cite{Dominguez:1989bz}. 
In the present work we concentrate on the thermal change of the lifetime of the $\eta'$ meson. As a follow-up work we also plan to determine for the $\eta'$ meson the thermal change of the coupling strength to the axial-vector current, i.e.\ how the overlap between the $\eta'$ and a quark current with the same quantum numbers --- the anomalous current --- varies as a function of the temperature. 

We do not focus on the transition region to chirally restored and deconfined matter but approach the problem from the 
low-temperature perspective.
In the intermediate-energy regime the physics of the strong interaction is governed by a plethora of hadronic resonance 
degrees of freedom without clear scale separations or large mass gaps. Here, systematic approaches like perturbation theory 
or effective field theories can only be applied to very carefully selected problems. This resonance dominated energy region 
corresponds to the 
temperature region where the transition to chirally restored and deconfined matter takes place \cite{Karsch:2003vd}. 
In general, one uses phenomenological models with hadronic and/or quark degrees of freedom to describe this energy/temperature 
regime \cite{Friman:2011zz}. Naturally such models cannot be systematically improved and are therefore accompanied 
by uncontrolled theory uncertainties.

In contrast, the low-energy regime can be described by Chiral Perturbation 
Theory ($ChPT$) \cite{Weinberg:1978kz,Gasser:1983yg,Gasser:1984gg,Scherer:2002tk}, an effective field theory (EFT) that utilizes the chiral symmetry and its spontaneous breakdown. Here, because of the 
Goldstone-boson nature of the lowest-energy excitations a significant mass gap exists that gives rise to scale separation, 
power counting and 
the possibility to systematically improve calculations and therefore to estimate the obtained accuracy of a prediction.  
The corresponding low-temperature region has been explored, e.g., 
in \cite{Gasser:1986vb,Gerber:1988tt,Goity:1989gs,Leutwyler:1990uq,Schenk:1991xe,Schenk:1993ru}. 
Actually the results from these works give some trust that one can address a temperature region that is not purely academic, 
i.e.\ temperatures that are reached in the late phase of relativistic heavy-ion collisions \cite{Friman:2011zz}. 
From the conceptual point of view one can study the onset of thermal modifications, i.e.\ the precursor effects for the 
transition happening higher up in temperature.

In the formal limit of an infinite number of quark colors $N_c$, the chiral anomaly vanishes and the $\eta'$ can then be included in the theory as the ninth Goldstone boson~\cite{Kaiser:2000gs,Witten:1980sp,DiVecchia:1980ve}. As a first step we calculate in the following the in-medium lifetime of the $\eta'$ meson based on Chiral Perturbation Theory for a large number of colors (large-$N_c \ ChPT$). We present a leading-order (LO) and a next-to-leading-order (NLO) calculation 
of the corresponding scattering amplitude. Formally this corresponds to an N$^5$LO calculation of the (imaginary part of the) 
self-energy of the $\eta'$ meson. 
While this is a conceptually well defined approach, there is one more 
issue to be considered when comparing to the real world. As long as $N_c$ is large enough there exists a mass gap 
between the then nine quasi-Goldstone bosons and all other excitations. In this case the only relevant degrees of freedom 
at low enough temperatures are the nine quasi-Goldstone bosons. In reality, however, there are other hadronic resonances with 
masses comparable to the physical mass of the $\eta'$ meson, notably scalar and vector mesons \cite{Agashe:2014kda}. These 
degrees of freedom should be taken into account if one addresses the in-medium properties of the $\eta'$ meson. Naturally 
this brings in some model uncertainties. To keep them as small as possible we follow the resonance-saturation approach of 
\cite{Ecker:1988te,Ecker:1989yg,Donoghue:1988ed}. Here hadronic resonances are introduced such that, once integrated out, 
they saturate the corresponding 
low-energy constants of NLO $ChPT$. Thus our second and more realistic approach to the 
in-medium properties of the 
$\eta'$ meson is based on a Lagrangian that includes (scalar and vector) 
resonances such that the model has the same formal low-energy, large-$N_c$ limit as $ChPT$. This is better known as Resonance Chiral Theory ($RChT$)~\cite{Pich:2008xj}.

Finally it is worth mentioning that the results of this work might be relevant also to heavy-ion collision physics. The  lifetime of the $\eta'$ in the vacuum~\cite{Agashe:2014kda} is about 600 times that of a ``fireball'' produced in heavy-ion collisions~\cite{Friman:2011zz,Tomasik:2005ng}. This means that, when studying the $\eta'$ in such experiments, one might miss the effects of the medium on the $\eta'$ if the meson lived much longer than the fireball. However at high temperatures and density we expect the lifetime of the $\eta'$ to become shorter due to interactions with particles (pions in our approximation) in the thermal medium.
If the lifetime of the $\eta'$ would become comparable with the lifetime of the fireball created in heavy-ion collisions, it would be possible to study experimentally the $\eta '$ and hence the change of some aspects of the axial anomaly at high temperature and density. The results of our studies support this possibility, i.e.\ show a considerable reduction in the lifetime of the $\eta'$ meson. For further studies of the $\eta'$ mesons in a medium see also~\cite{Vertesi:2009wf,Itahashi:2012ut,Nanova:2013fxl,Nanova:2012vw,Sakai:2013nba,Kwon:2012vb,Jido:2011pq}.

The present paper is organized as follows. In Section~\ref{Sec2} we present the methods used to estimate the changes in the width of the $\eta'$ meson at low temperature. In Section~\ref{Sec3} we introduce the two frameworks in which the calculations are performed: $ChPT$ and $RChT$. In particular we study the behavior of the $\eta'$ width in both frameworks and we discuss the corresponding results. Finally in Section~\ref{Sec4} we summarize the content of this article and in Section~\ref{Sec5} we give some suggestions for future research.

\section{
The thermal width of the $\eta'$}
\label{Sec2}
The purpose of this work is to calculate the thermal width of the $\eta'$, where the medium is a hadronic gas. All the used methods and approximations are illustrated below. Note that we do not consider any non-vanishing chemical potential. The complementary case of cold nuclear matter is studied experimentally in~\cite{Itahashi:2012ut,Nanova:2013fxl,Nanova:2012vw} and within various models in~\cite{Oset:2010ub,Sakai:2013nba,Jido:2011pq}.
\subsection*{Width increase at nonzero temperature}
When we place a bosonic particle in a medium, we expect an increase of its width due to mainly two factors, the Bose enhancement and the collisional broadening:
\begin{equation*}
\Gamma_{T\neq0}=\Gamma_{BE}+\Delta\Gamma_{\mathrm{coll}},
\end{equation*}
where the Bose enhanced decay width $\Gamma_{BE}$ takes also into account the ordinary vacuum decay width. In our case of interest the contribution given by $\Gamma_{BE}$ is negligible if compared with that of $\Delta\Gamma_{\mathrm{coll}}$~\cite{Reference2}, and we are therefore allowed to make a further approximation:
\begin{equation*}
\Gamma_{T\neq0}\approx\Delta\Gamma_{\mathrm{coll}}.
\end{equation*}
In the present work we concentrate on an $\eta'$ meson at rest with respect to the thermal medium. For a heavy-ion collision, where the fireball extends only over a limited volume, we expect that thermal modifications are most pronounced for probes that are at rest with respect to this thermal surrounding.

In the low-density approximation the collisional broadening is given by~\cite{Leupold:2009kz,Reference2,Das:1997gg}
\begin{equation}\label{Gamma_coll}
\Delta\Gamma_{\mathrm{coll}}=\int{\frac{\mathrm{d}^3p}{(2\pi)^3}\sum_{i}n_{B,F}(E_{ip})\frac{\vert\bar{p}\vert}{E_{ip}}\sum_{X} \sigma_{\eta'i\rightarrow X}(E_{ip})},
\end{equation}
where $\sigma_{\eta'i\rightarrow X}(E_{ip})$ is the cross section for inelastic scattering of an $\eta'$ at rest and a medium particle $i$ with momentum $\bar{p}$ and energy $E_{ip}$. We have to sum over all types of heat bath particles and over all possible final states $X$. The particles forming the medium follow a Bose-Einstein or Fermi-Dirac distribution: $$n_{B,F}(E_{ip})=\frac{1}{e^{E_{ip}/T}\mp1}.$$
We need to use several approximations in order to simplify the calculations. First of all the medium, consisting of a hadronic gas, will be simulated by a gas of pions, the lightest hadrons. This is a very good approximation at low temperature, where heavier particles with mass $M$ are essentially suppressed as $e^{-M/T}$ by the thermal distribution~\cite{Leutwyler:1990uq,Gerber:1988tt,Gasser:1986vb,Goity:1989gs,Schenk:1993ru,Schenk:1991xe}.
The collisional broadening can be rewritten from Eq.~\eqref{Gamma_coll} as
\begin{equation}\label{Gamma_collnew}
\Delta\Gamma_{\mathrm{coll}}=\frac{1}{(2\pi)^2}\int{\mathrm{d}E_p\left(E^2_p-M^2_{\pi}\right)n_B(E_p)\sum_{i,X} \sigma_{\eta'\pi_i\rightarrow X}(E_p)},
\end{equation}
where we need to sum over all the inelastic cross sections involving an $\eta'$ and any type of pion in the initial state.
Which are then the relevant processes?
Since in our approximation the $\eta'$ can only interact with pions, we will consider two types of processes that contribute to collisional broadening: $\eta'\pi\rightarrow\eta\pi$ and $\eta'\pi\rightarrow\bar{K}K$. We work in the isospin limit and therefore disregard the isospin violating reaction $\eta' \pi\rightarrow\pi\pi$. We also disregard reactions where more than two particles are produced since such cross sections are suppressed by phase space and in $ChPT$. 
Having defined isospin multiplets according to: 
$$\pi=
\left(
\begin{array}{c}
\pi^+ \\
\pi^0 \\
\pi^-
\end{array}
\right), \qquad
K=
\left(
\begin{array}{c}
K^+ \\
K^0 
\end{array}
\right), \qquad
\bar{K}=
\left(
\begin{array}{c}
\bar{K}^{0} \\
K^-
\end{array}
\right),
$$
we only have to calculate $\eta'\pi\rightarrow\eta\pi$ and $\eta'\pi\rightarrow K\bar{K}$, where $\pi$ indicates one of the fields in the triplet, $K$ and $\bar{K}$ one of the fields in the doublet. Since there are only two possible final states for processes with a charged pion  $\eta'\pi^{\pm}\rightarrow\eta\pi^{\pm}/K^0K^{\pm}$ (there are three with the neutral pion), we choose to calculate the cross sections that involve one of the charged pions for simplicity. 
Assuming exact isospin symmetry, the corresponding matrix elements are the same. In addition it follows that the masses of the charged pions are identical to that of the neutral pion. This implies that we will obtain the same value of the cross section:
$$\sigma(\eta'\pi^+\rightarrow\eta\pi^+)=\sigma(\eta'\pi^-\rightarrow\eta\pi^-)=\sigma(\eta'\pi^0\rightarrow\eta\pi^0)\,.$$
The same reasoning holds also for the case with two kaons in the final state. Once more note that the process with a neutral pion in the initial state gives rise to two different kaon final states and therefore would require more calculations. We have:
\begin{equation*}\begin{aligned}
\sigma(\eta'\pi^+\rightarrow K^+ \bar{K}^0)&=\sigma(\eta'\pi^-\rightarrow K^- K^0)\\ &=\sigma(\eta'\pi^0\rightarrow K^+ K^-)+ \sigma(\eta'\pi^0\rightarrow K^0\bar{K}^0)\,.
\end{aligned}\end{equation*}
The formula for $\Delta\Gamma_{\mathrm{coll}}$~\eqref{Gamma_collnew} contains the sum of the cross sections for all the possible processes arising from the collision of a pion and the $\eta'$, i.e.
\begin{equation*}
\sum_{i}\sigma_i=3[\sigma(\eta'\pi^+\rightarrow\eta\pi^+)+\sigma(\eta'\pi^+\rightarrow K^+ \bar{K}^0)],
\end{equation*}
where we have chosen the positive-pion processes for convenience.
We simply have to calculate these two cross sections, sum them up and multiply by three since for any pion in the isospin triplet the cross section is the same.

The collisional broadening gets two separate contributions: $\Delta\Gamma_{\eta\pi}$ from the $\eta\pi$ final state processes and $\Delta\Gamma_{K\bar{K}}$ from the processes with $K\bar{K}$ in the final state. Therefore the total increase of the width is approximately given by:
\begin{equation*}
\Delta\Gamma_{\mathrm{coll}}\approx\Delta\Gamma_{\eta\pi}+\Delta\Gamma_{K\bar{K}}.
\end{equation*}
In the next section we will determine the two cross sections needed to calculate the collisional width.

\section{Formalism and results}
\label{Sec3}

\subsection{Large-$N_{c}$ Chiral Perturbation Theory}
\subsubsection{Formalism}
The advantage of working in the framework of large-$N_c\ ChPT$ is that one can include in the theory a ninth Goldstone boson: the $\eta'$~\cite{Kaiser:2000gs}.
When $N_c\rightarrow\infty$ the chiral symmetry is promoted to $G=U(3)_L\times U(3)_R$. Since the vacuum is symmetric with respect to the subgroup $H=SU(3)_V\times U(1)_V$, the Goldstone bosons are described by the coset space $G/H\simeq U(3)$.
This implies that the meson fields should be parametrized by a unitary matrix $U(x)\in U(3)$, such as~\cite{DiVecchia:1980sq, Witten:1980sp, DiVecchia:1980ve}:
\begin{equation}\label{U}
U=\mathrm{e}^{i\phi}
\end{equation}
where the $\phi$ matrix is in terms of the bare meson fields 
\begin{equation*}
\phi\equiv
\begin{pmatrix}
\pi^0_B+\frac{1}{\sqrt{3}}\eta_8+\sqrt{\frac{2}{3}}\eta_0 & \sqrt{2}\pi^+_B & \sqrt{2}K^+_B \\
\sqrt{2}\pi^-_B & -\pi^0_B+\frac{1}{\sqrt{3}}\eta_8+\sqrt{\frac{2}{3}}\eta_0 & \sqrt{2}K^0_B \\ \sqrt{2}K^-_B & \sqrt{2}\bar{K}^0_B & -\frac{2}{\sqrt{3}}\eta_8+\sqrt{\frac{2}{3}}\eta_0
\end{pmatrix}.
\end{equation*}
We will see later how these bare fields are related to the physical fields. For example, linear combinations of the $\eta_0$ and the $\eta_8$ fields give rise to the physical $\eta$ and $\eta'$ mesons. Note that $U(x)$ can be decomposed into:
\begin{equation*}
U=\mathrm{e}^{i\sqrt{2/3}\eta_0} \tilde{U},
\end{equation*}
where $\tilde{U}\in SU(3)$ is multiplied by a phase factor. With respect to ordinary $SU(3)\ ChPT$~\cite{Gasser:1984gg} we have gained an additional degree of freedom:
 the determinant of the $U(3)$ matrix, which describes the ninth Goldstone boson. 

\subsection*{Effective Lagrangian at LO and NLO}
The expansion of the effective Lagrangian goes as
\begin{equation}\label{effL}
\mathcal{L}_{\mathrm{eff}}=\sum_{i=0}^{\infty}\mathcal{L}^{(i)},
\end{equation}
where according to~\cite{Kaiser:2000gs} we use a single power-counting parameter $\delta$ 
\begin{equation*}
p^2=O(\delta),\quad m_q=O(\delta),\quad 1/N_c=O(\delta)
\end{equation*}
with $p$ denoting a typical momentum and $m_q$ a quark mass. In this notation the contributions to the effective Lagrangian from $\mathcal{L}^{(i)}$ are of order $O(\delta^i)$.
The LO large-$N_c \ ChPT$ Lagrangian, invariant with respect to local chiral transformations, is order $O(\delta^0)=O(1)$. It reads~\cite{Kaiser:2000gs}:
\begin{equation}\label{LO}
\mathcal{L}^{(0)}=\frac{1}{4}F^2\mathrm{Tr}~[D_\mu U^{\dagger}D^\mu U]+\frac{1}{4}F^2\mathrm{Tr}~[U\chi^{\dagger}+\chi U^{\dagger}]-3\tau(\eta_0)^2,
\end{equation}
where $D_\mu$ is the chiral gauge covariant derivative and $\chi =2B(s+ip) \rightarrow 2B\mathcal{M}$ with $\mathcal{M}=\mathrm{diag}(m,m,m_s)$ the quark mass matrix in the isospin limit. The low-energy constants $F,B$ and $\tau$ will be specified below.
Since the singlet field $\eta_0$ is included in $\phi$ and therefore also in $U$, it follows that the $\eta_0$ mass will get a contribution from the nonzero quark masses and another from the topological susceptibility $\tau$. This reminds us that the $U(1)_A$ symmetry is not only broken by the quark masses but also by the anomaly.
Note that~\eqref{LO} is order $O(1)$ since the decay constant $F$ depends on $N_c$ as  $F=O(\sqrt{N_c})$ so that $F^2=O(1/\delta)$ and $\tau$ is $O(1)$.

Putting the source $s$ equal to the quark mass matrix and all the other external fields to zero (so that $\chi=2B\mathcal{M}$), the NLO Lagrangian reads~\cite{Kaiser:2000gs}:
\begin{equation}\begin{split}\label{NLO}
\mathcal{L}^{(1)}&=L_2\mathrm{Tr}\big[\partial_\mu U^{\dagger}\partial_\nu U \partial^\mu U^{\dagger}\partial^\nu U\big]+(2L_2+L_3)\mathrm{Tr}\big[\partial_\mu U^{\dagger} \partial^\mu U\partial_\nu U^{\dagger}\partial^\nu U\big]\\&\phantom{=}+2BL_5\mathrm{Tr}\big[\partial_\mu U^{\dagger} \partial^\mu U(U^{\dagger}\mathcal{M}+\mathcal{M}U)\big]\\&\phantom{=}+4B^2L_8\mathrm{Tr}\big[U^\dagger\mathcal{M}U^\dagger\mathcal{M}+\mathcal{M}U\mathcal{M}U\big]\\&\phantom{=}+\frac{1}{2}F^2\Lambda_1\partial_\mu\eta_0\partial^\mu\eta_0-\frac{i}{\sqrt{6}}BF^2\Lambda_2\eta_0\mathrm{Tr}\big[U^\dagger\mathcal{M}-\mathcal{M}U\big],
\end{split}\end{equation}
with the low-energy constants $L_i\sim N_c$ and $\Lambda_i\sim 1/N_c$.
This Lagrangian is of order $O(\delta)$. 
The large-$N_c$ limit presents another great advantage with respect to the $N_c=3$ case: working at $O(\delta)$ one has to include only tree level diagrams from $\mathcal{L}^{(0)}$ and $\mathcal{L}^{(1)}$. This is a consequence of the fact that meson loops are suppressed in the $1/N_c$ expansion~\cite{Donoghue:1992dd,Kaiser:2000gs,'tHooft:1974hx,'tHooft:1973jz} and therefore moved one step up in the power counting. This means that loop diagrams with vertices from $\mathcal{L}^{(0)}$ appear first at $O(1/N_c^2)=O(\delta^2)$, and therefore we will not consider them in this work.

\subsection*{Relating bare and physical fields}\label{relations}
We need a relation between the bare fields in terms of which the effective Lagrangian is given and the physical fields~\cite{Gasser:1984gg}.
Let us define nine-dimensional column vectors that contain all the meson fields. The one collecting bare fields is $\bar{\phi}_B$, while the physical fields form $\bar{\phi}_P$.
The relation between them is:
\begin{equation}
\bar{\phi}_P=\textbf{F}\bar{\phi}_B,
\end{equation}
where $\textbf{F}$ is the following block-diagonal matrix:
$$
\textbf{F}=\left(
\begin{array}{c|c}
  \textbf{F}_{\pi K} & 0 \\
  \hline
  0 & \textbf{F}_{\eta} \ \ 
 \end{array}
 \right).$$
The matrix $\textbf{F}_{\pi K}$ is also diagonal. The reason is that in our framework we have exact conservation of isospin (and strangeness). Therefore the only states that can mix are $\eta_0$ and $\eta_8$. The elements on the diagonal of $\textbf{F}_{\pi K}$ are the pion and the kaon decay constant, $F_\pi$ and $F_K$, respectively. On the other hand $\textbf{F}_\eta$ is not diagonal and it is given in terms of two mixing angles and two decay constants~\cite{Leutwyler:1997yr,Kaiser:1998ds}:
\begin{equation*}
\textbf{F}_\eta=
\begin{pmatrix}
F_8\cos\theta_8 & -F_0\sin\theta_0 \\
F_8\sin\theta_8 & F_0\cos\theta_0
\end{pmatrix}.
\end{equation*}
Now it is immediate to express the bare fields in terms of the physical ones. Regarding the pion and kaon fields, the bare fields are equal to the physical fields divided by the correspondent decay constant:
\begin{equation*}\label{PK}
\pi_B=\frac{\pi_P}{F_\pi} \qquad\text{and}\qquad K_B=\frac{K_P}{F_K}\,.
\end{equation*}
The bare fields $\eta_0$ and $\eta_8$ are related to the physical fields $\eta$ and $\eta'$ by the inverse of the mixing matrix, $\textbf{F}^{-1}_\eta$
\begin{equation*}\label{etas}
\begin{pmatrix}
\eta_8\\ \eta_0
\end{pmatrix} =
\frac{1}{F_8F_0\cos(\theta_8-\theta_0)}
\begin{pmatrix}
F_0\cos\theta_0&F_0\sin\theta_0\\
-F_8\sin\theta_8&F_8\cos\theta_8\\
\end{pmatrix}
\begin{pmatrix}
\eta\\ \eta'
\end{pmatrix}.
\end{equation*}
Note that the decay constants here play the role of a wave function renormalization and provide the physical fields with the right mass dimension.

\subsection*{Determination of the LEC:s}\label{determinationLEC}
The coupling constants in the effective Lagrangian are referred to as low-energy constants (LEC:s). We will briefly summarize the procedure to determine them, more details can be found in~\cite{Reference2,Leutwyler:1997yr}.

As said before, the effective Lagrangian is given in terms of the bare fields. However we know that the free Lagrangian in terms of the physical fields has the following form:
\begin{equation}\label{Lfree}
\mathcal{L}_{\mathrm{eff,free}}=\sum_P \partial^\mu\phi_P\partial^\mu\phi_P-\sum_PM_P^2\phi^2_P\,.
\end{equation}
In order to compare this with the effective Lagrangian~\eqref{effL} we need to express the physical fields in terms of the bare ones. 
Instead of using mixing angles it is convenient to introduce the following four new parameters:
\begin{equation*}\begin{aligned}
F_\eta^8&=F_8\cos\theta_8, & F_\eta^0&=F_0\sin\theta_0, \\
F_{\eta'}^8&=F_8\sin\theta_8, & F_{\eta'}^0&=F_0\cos\theta_0.
\end{aligned}\end{equation*}
By rewriting Eq.~\eqref{Lfree} in terms of the bare fields we obtain:
\begin{equation}\begin{split}
\mathcal{L}_{\mathrm{eff,free}}&=F^2_\pi\left(\frac{1}{2}\partial^\mu\pi_B^0\partial^\mu\pi_B^0+\partial^\mu\pi_B^+\partial^\mu\pi_B^-\right)-M_\pi^2F^2_\pi\left[\frac{1}{2}(\pi^0_B)^2+\pi^+_B\pi^-_B\right] \\&\phantom{=} + F^2_K\left(\partial^\mu K_B^0\partial^\mu\bar{K}^0_B+\partial^\mu K_B^+\partial^\mu K_B^-\right)-M_K^2F^2_K\left(K^0_B \bar{K}^0_B+K^+_B K_B^-\right) \\& \phantom{=}+\frac{1}{2}\left[(F_\eta^8)^2+(F_{\eta'}^8)^2\right]\partial_\mu\eta_8\partial^\mu\eta_8-\frac{1}{2}\left[(F_\eta^8)^2M_\eta^2+(F_{\eta'}^8)^2M_{\eta'}^2\right](\eta_8)^2 \\& \phantom{=}+\frac{1}{2}\left[(F_\eta^0)^2+(F_{\eta'}^0)^2\right]\partial_\mu\eta_0\partial^\mu\eta_0-\frac{1}{2}\left[(F_\eta^0)^2M_\eta^2+(F_{\eta'}^0)^2M_{\eta'}^2\right](\eta_0)^2 \\&\phantom{=} +\left(F_\eta^8F_\eta^0+F_{\eta'}^8F_{\eta'}^0\right)\partial_\mu\eta_8\partial^\mu\eta_0-\left(F_\eta^8F_\eta^0M_\eta^2+F_{\eta'}^8F_{\eta'}^0M_{\eta'}^2\right)\eta_8\eta_0.
\end{split}\end{equation} 
Once we have expanded the traces in the free part of $\mathcal{L}^{(0)}+\mathcal{L}^{(1)}$, keeping all terms of second order in $\phi$, we can compare term by term the coefficients in front of the bare fields with the expression above, which is already written in terms of the bare fields. In doing so, we will obtain ten equations in terms of the unknown LEC:s. The full set of relations is:
\begin{equation}\begin{aligned}\label{system}
&F^2_\pi=F^2+16BmL_5, \\
&F^2_K=F^2+8BmL_5(1+S), \\
&3F^2_8=3F^2+16BmL_5(1+2S), \\
&3F^2_0=3F^2+3\Lambda_1F^2+16BmL_5(2+S), \\
&3 F_{0}F_{8} \sin(\theta_{8}-\theta_{0}) = 16 \sqrt{2} B m L_{5} (1-S), \\
&F^2_\pi M^2_\pi =2F^2Bm+64(Bm)^2L_8, \\
&F^2_K M^2_K =F^2Bm(1+S)+16(Bm)^2L_8(1+S)^2, \\
&3F^2_8\left(M_\eta^2\cos^2\theta_8+M_{\eta'}^2\sin^2\theta_8\right)=2F^2Bm(1+2S)+64(Bm)^2L_8(1+2S^2), \\
&3F^2_0\left(M_\eta^2\sin^2\theta_0+M_{\eta'}^2\cos^2\theta_0\right)=2F^2Bm(2+S)+4F^2Bm\Lambda_2(2+S) \\
& \hspace{14em} +64(Bm)^2L_8(2+S^2)+18\tau, \\
&3 F_{0}F_{8}\left(-M^{2}_{\eta}\sin\theta_{0}\cos\theta_{8}+M^{2}_{\eta '}\cos\theta_{0}\sin\theta_{8}\right)= \\ 
& \hspace{5em} 2\sqrt{2}\left[F^{2}Bm(1-S)+F^{2}Bm\Lambda_{2}(1-S)+32(Bm)^{2}L_{8}(1-S^{2})\right].
\end{aligned}\end{equation}
These relations are valid to $O(\delta)$. We need them to determine the unknown LEC:s, which we take to be the following twelve parameters:
\begin{equation}\label{parameters}
F,\ S,\ Bm,\ \theta_0,\  \theta_8,\ F_0,\ F_8,\ L_5,\ L_8,\ \Lambda_1,\ \Lambda_2\ \text{and } \tau.
\end{equation}
Note that $Bm$ is taken as a single parameter and not two. This is because we are not able to isolate the  quark masses; we can only determine $B$ times the magnitude of the quark masses, along with the quark mass ratio $S\equiv m_s/m$~\cite{Gasser:1984gg,Gasser:1983yg}.
We use the remaining parameters as input:
\begin{equation}\begin{aligned}\label{input}
M_{\pi}&= 138.04 \text{ MeV}, & M_{K}&= 494.99 \text{ MeV}, \\
M_{\eta}&= 547.86 \text{ MeV},& M_{\eta'}&= 957.78\text{ MeV},\\
F_\pi&= 92.21\text{ MeV}, & F_K&= 110.47\text{ MeV}.
\end{aligned}\end{equation}
The meson masses\footnote{$M_\pi$ and $M_K$ are obtained as the average of the masses in the respective isospin multiplets.} and the values of the decay constants come from~\cite{Agashe:2014kda}. 
We now have a system of ten equations~\eqref{system} and twelve parameters~\eqref{parameters}. We still need two more equations to be able to solve it. In accordance with~\cite{Leutwyler:1997yr} we exploit the electromagnetic interactions of the $\eta$ and the $\eta'$ to obtain two more relations. In particular we look at the anomalous decays $\eta\rightarrow\gamma\gamma$ and $\eta'\rightarrow\gamma\gamma$. Then we need to add the Wess-Zumino-Witten term~\cite{Wess:1971yu,Witten:1983tw} to the effective Lagrangian. The piece relevant for the above transitions is given by:
\begin{equation}\label{Lwzw}
\mathcal{L}_{WZW}=-\frac{N_c\alpha}{4\pi}F_{\mu\nu}\tilde{F}^{\mu\nu}\mathrm{Tr}[\mathcal{Q}^2\phi]
\end{equation}
where $F_{\mu\nu}$ is the electromagnetic field strength tensor, $\alpha$ the electromagnetic fine-structure constant and $\mathcal{Q}$=diag(2/3,-1/3,-1/3) is the quark-charge matrix. Since the Wess-Zumino-Witten term is not invariant under a change of scale beyond contributions of order $O(\delta)=O(1/N_c)$, the term below (which belongs to the $\mathrm{N^2LO}$ Lagrangian) is needed in order to get rid of the scale dependence~\cite{Kaiser:2000gs,Leutwyler:1997yr}
\begin{equation}
\mathcal{L}_{WZW}^{(2)}=-\frac{N_c\alpha\Lambda_3}{18\pi}F_{\mu\nu}\tilde{F}^{\mu\nu}\sqrt{6}\eta_0\,.
\end{equation}
From~\eqref{Lwzw} we derive the following decay rate:
\begin{equation}\label{decay}
\Gamma_{P\rightarrow\gamma\gamma}=\frac{\alpha^2N_c^2}{576\pi^3F_\pi^2}M^3_Pc^2_P
\end{equation}
with $c_P=1$ for the pion, i.e. for \ $P=\pi$. Thanks to the expression~\eqref{decay} we can estimate the constant $c_P$ for $P=\eta,\ \eta'$ from experimental data for the decay width of the $\eta$ and $\eta'$ mesons~\cite{Agashe:2014kda}. We obtain the values $c_\eta=0.997$ and $c_{\eta'}=1.253$.
Finally we would like to find a way to link them with some of the unknown LEC:s. When we consider the mixing between the $\eta$ and the $\eta'$, we get relations between the photonic decay rates. We have then:
\begin{equation}\label{extra}\begin{aligned}
&\sqrt{3}\left(F^8_\eta c_\eta+F^8_{\eta'} c_{\eta'}\right)=\sqrt{3}F_8(c_\eta\cos\theta_8+c_{\eta'}\sin\theta_8)=F_\pi,\\
&\sqrt{3}\left(F^0_\eta c_\eta+F^0_{\eta'} c_{\eta'}\right)=\sqrt{3}F_0(-c_\eta\sin\theta_0+c_{\eta'}\cos\theta_0)=\sqrt{8}F_\pi(1+\Lambda_3).
\end{aligned}\end{equation}
These two relations together with~\eqref{system} allow us to solve a system of twelve equations with twelve unknown variables. Note that the factor $(1+\Lambda_3)$ in~\eqref{extra} comes from the renormalization term mentioned above. This implies that some of the LEC:s have a scale dependence, but as we show in \ref{A} they do not affect the observable quantities of our interest (scattering amplitudes) since they show up only in scale invariant combinations.  
\subsection*{Numerical values of the LEC:s}
Following~\cite{Leutwyler:1997yr} and using as input the masses $M_\pi,\ M_K, \ M_\eta \text{ and } M_{\eta'}$ and the two decay constants $F_\pi$ and $F_K$ with values according to~\eqref{input}, we find the following numerical values (for details see also~\cite{Reference2}):
\begin{equation}
\begin{gathered}
F= 90.4\text{ MeV},\quad Bm=(97.23\text{ MeV})^2,\quad S=23.3, \\
L_5=2.19\cdot10^{-3},\quad L_8=1.31\cdot10^{-3},\\
F_8=115.9 \text{ MeV},\quad \theta_{8}=-21.9^\circ,\\  
\frac{F_0}{1+\Lambda_3}=110.5 \text{ MeV},\quad \theta_{0}=-6.8^\circ.
\end{gathered}\label{values}
\end{equation}
Note that in the solution for $F_0$ the scale dependent parameter $\Lambda_3$ appears. Also $\Lambda_1,\ \Lambda_2$ and $\tau$ depend on the renormalization scale and thus are not observable quantities. However working to $O(\delta)$ we find the three scale invariant combinations:
\begin{equation}
\Lambda_1-2\Lambda_3=0.152, \quad \Lambda_2-\Lambda_3=0.228, \quad \frac{\tau}{(1+\Lambda_3)^2}=(190\text{ MeV})^4.
\end{equation}
We will explicitly see that the matrix elements for the scattering amplitudes contain only these scale invariant combinations.
As one might have already noticed, two other LEC:s appear in the effective Lagrangian~\eqref{NLO}: $L_2$ and $L_3$. We will be able to find their numerical values using relations between them and some $RChT$ parameters that will be introduced later.

\subsubsection{Results}
\begin{figure}[h]
 \centering
   \includegraphics[width=8cm]{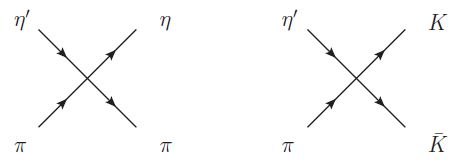}
   \caption{The $ChPT$ diagrams contributing to the in-medium width of the $\eta^\prime$ .}\label{ChPT}
 \end{figure}

In order to calculate the cross sections that enter the formula for the collisional broadening~\eqref{Gamma_collnew}, we derive the scattering amplitudes for the processes $\eta'\pi\rightarrow\eta\pi$ and $\eta'\pi\rightarrow K\bar{K}$ from the effective Lagrangian~\eqref{effL}. The matrix elements can be found in \ref{A} while the corresponding four-point vertex diagrams are shown in Fig.~\ref{ChPT}. For more details on their derivation we refer to~\cite{Reference2}. From a formal point of view the width of a state is related to the imaginary part of a self-energy~\cite{Das:1997gg,Leupold:2009kz}. The imaginary part is obtained from the optical theorem, which corresponds on the digrammatic level to cutting the propagator lines.
With the diagrams from Fig.~\ref{ChPT} the corresponding self-energy is a two-loop ``sunset'' diagram. Using vertices from the LO Lagrangian~\eqref{LO} of large-$N_c \ ChPT$, the corresponding two-loop diagram is of formal order $\delta^5$, i.e. an $\mathrm{N^4LO}$ contribution (evaluated in our work in the linear-density approximation). Using one vertex from the NLO Lagrangian~\eqref{NLO} yields the $\mathrm{N^5LO}$ contribution. Here our accuracy is terminated since also not considered three-loop diagrams would contribute at $\mathrm{N^6LO}$. While $\mathrm{N^5LO}$ sounds impressive we would like to stress that we calculate only the imaginary part of the self-energy. The LO and NLO contributions to two-point functions do not contain loops. $\mathrm{N^2LO}$ and $\mathrm{N^3LO}$ self-energies are just ``snail'' diagrams that have no imaginary part. Instead of calling our results for the collisional width $\mathrm{N^4LO}$ and $\mathrm{N^5LO}$, we have decided to label the results according to the formal order of the respective matrix elements LO and NLO.

\begin{figure}[h]
	\centering
		\includegraphics[width=10cm]{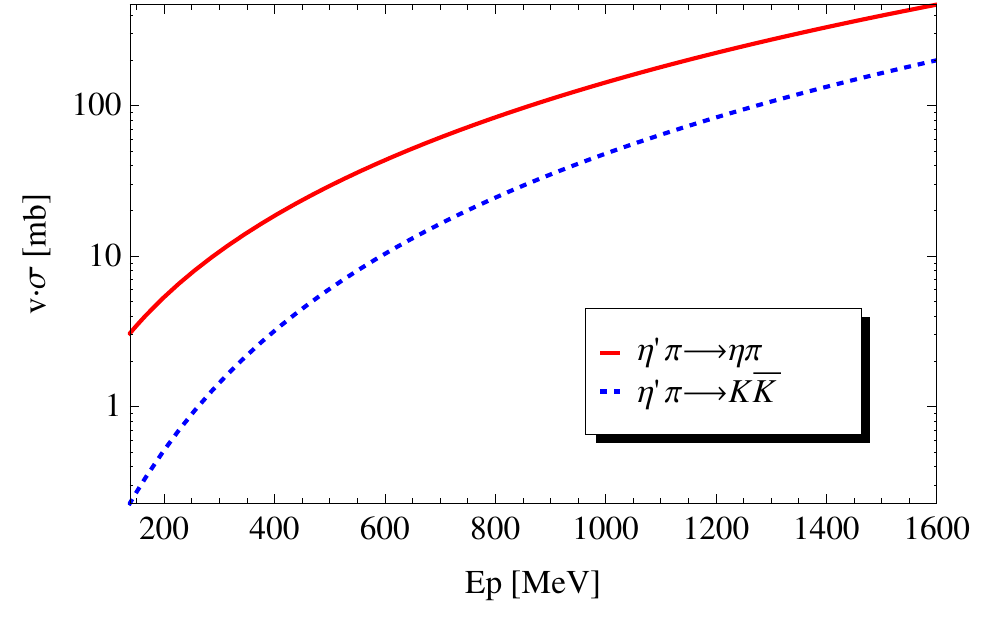}
	\caption{The reaction rates $v\cdot\sigma$ for the processes $\eta' \pi\rightarrow\eta\pi$ and $\eta' \pi\rightarrow\bar{K}K$ at NLO in ChPT, plotted as functions of the incoming pion energy $E_p\ge M_\pi$ in the rest frame of the $\eta'$. These quantities enter the collisional width integral.}
	\label{rateChPT}
\end{figure}

We start presenting the NLO large-$N_c \ ChPT$ results for the width increase of the $\eta'$, which will be later compared with the $RChT$ results. 
Note that for our work the quantity of physical interest is the reaction rate, not directly the cross section. This is in fact the quantity that enters the collisional width~\eqref{Gamma_coll}. If the volume of the system is normalized to 1, then the reaction rate is $v_{rel}\times\sigma$ where $v_{rel}=\left|\bar{p}\right|/E_p$ is the velocity of the incoming pion in the rest frame of the $\eta'$. Fig.~\ref{rateChPT} shows the reaction rates of both the processes $\eta' \pi\rightarrow \eta\pi$ and $\eta' \pi\rightarrow K\bar{K}$, plotted as functions of the incoming pion energy $E_{p}$ in the rest frame of the $\eta'$. 
In the framework of $ChPT$ we obtain the matrix elements $\mathcal{M}_{\eta' \pi\rightarrow \eta\pi}^{ChPT}$ and $\mathcal{M}_{\eta' \pi\rightarrow \bar{K}K}^{ChPT}$ which have an energy dependence at NLO (see~\eqref{etapi},~\eqref{kaons}). Even at LO the kaonic matrix element has an energy dependence while $\mathcal{M}_{\eta' \pi\rightarrow \eta\pi}^{ChPT}$ is a constant (see~\eqref{LOkaons},~\eqref{LOetapi}). It turns out that at LO (not shown here) the cross section with kaonic final states is much larger than the one with an $\eta$ and a pion in the final state~\cite{Reference2}, i.e.\ it is opposite to the NLO result. This is an artefact of the LO calculation that is lifted at NLO; see also the discussion in~\cite{Escribano:2000ju,Escribano:2010wt}. Getting back to  NLO we can see from Fig.~\ref{rateChPT} that both the reaction rates continue to rise for higher pion energies $E_{p}$, due to the energy dependence of the matrix elements. The $\eta \pi$ final state has a higher reaction rate than the  $\bar{K}K$ final state from which it follows that the collisional broadening due to $\eta' \pi\rightarrow \eta\pi$ reactions is much larger than that from $\eta' \pi\rightarrow \bar{K}K$ reactions. 
The sizes of the two reaction rates vary from around 1-10 mb at $E_{p}\approx200$ MeV up to about 400 mb (for final state $\eta\pi$) at $E_{p}\approx1500$ MeV. In particular, it reaches rather unrealistic values for $E_{p}>800$ MeV. How can we interpret this rising of the cross section in relation with high pion energies?
A plausible explanation is that this tendency to grow is a signal of $ChPT$ break down. In fact, already at energies $\approx1$ GeV, effects ignored in $ChPT$ start to be important, and we can no longer trust $ChPT$ calculations. Among these effects, we must mention those generated from resonance exchange. It is the aim of $RChT$ to include these effects. We will address these issues below.

\begin{figure}[h]
	\centering
		\includegraphics[width=11cm]{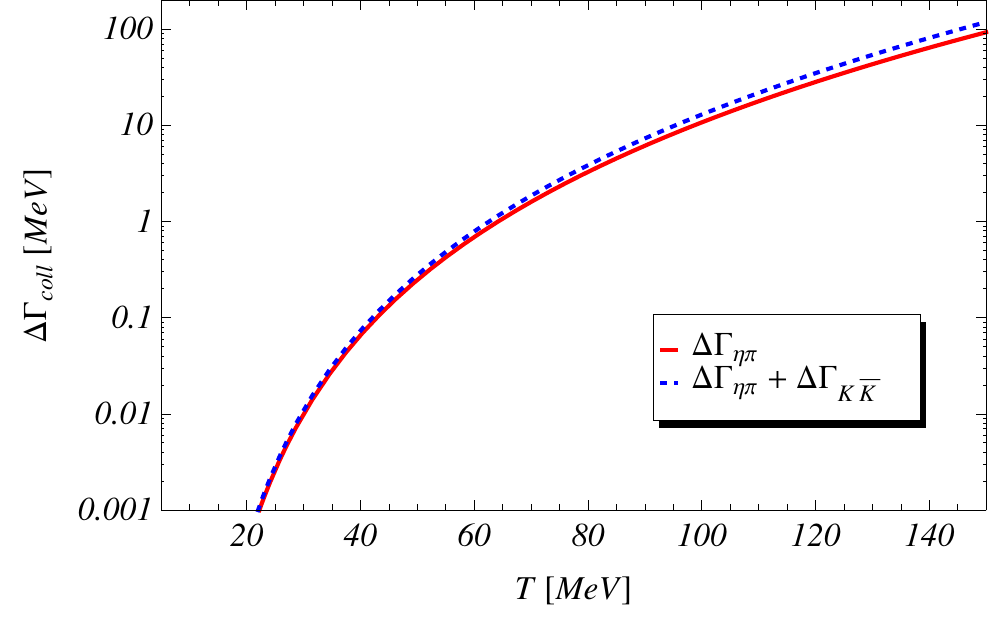}
	\caption{Addition to the $\eta'$ width as a function of the temperature at NLO in large-$N_c \ ChPT$.}
	\label{fig:Getapi}
\end{figure}
Continuing with the $ChPT$ results, the increase of the $\eta'$ width in a pion gas is shown in Fig.~\ref{fig:Getapi}.
These first results already allow us to make some comparisons. Let us assume that the lifetime of a fireball is $\tau_{fb}\approx10-20$ fm/c~\cite{Friman:2011zz,Tomasik:2005ng}. This means that the in-medium lifetime of the $\eta'$ would become comparable with that of a fireball if its width would increase by $\Delta\Gamma_{\mathrm{coll}}\approx20$ MeV, i.e.\ $1/\tau_{fb}$. Recall that the width of the $\eta'$ in the vacuum is very small, $\Gamma_{vac}^{\eta'}\approx 200$ keV~\cite{Agashe:2014kda}. We can deduce from Fig.~\ref{fig:Getapi} that this seems to happen for temperatures around $T\approx100$ MeV. The width increases considerably in the medium, causing therefore significant changes in the lifetime. All this would apply, if we could trust the pure $ChPT$ results.

In order to judge the reliability of the results for the width, we need to focus our attention on the integrand in the collisional broadening integral in Eq.~\eqref{Gamma_collnew}:
\begin{equation}\label{integrand}
f(E_{p},T)= \left(E_{p}^{2}-M_{\pi}^2\right)n_{B}(E_{p},T)\left[\sigma_{\eta\pi}(E_{p})+\sigma_{K\bar{K}}(E_{p})\right],
\end{equation}
where $n_{B}(E_{p},T)$ is the Bose-Einstein distribution and $\sigma_{\Phi_a\Phi_b}(E_{p})$ denotes the cross section $\sigma(\eta' \pi\rightarrow \Phi_a\Phi_b)$. If the dominant contribution to the integral stems from low energies where we might trust $ChPT$, then we might also trust the results of Fig.~\ref{fig:Getapi}. 
\begin{figure}
	\centering
		\includegraphics[width=\textwidth]{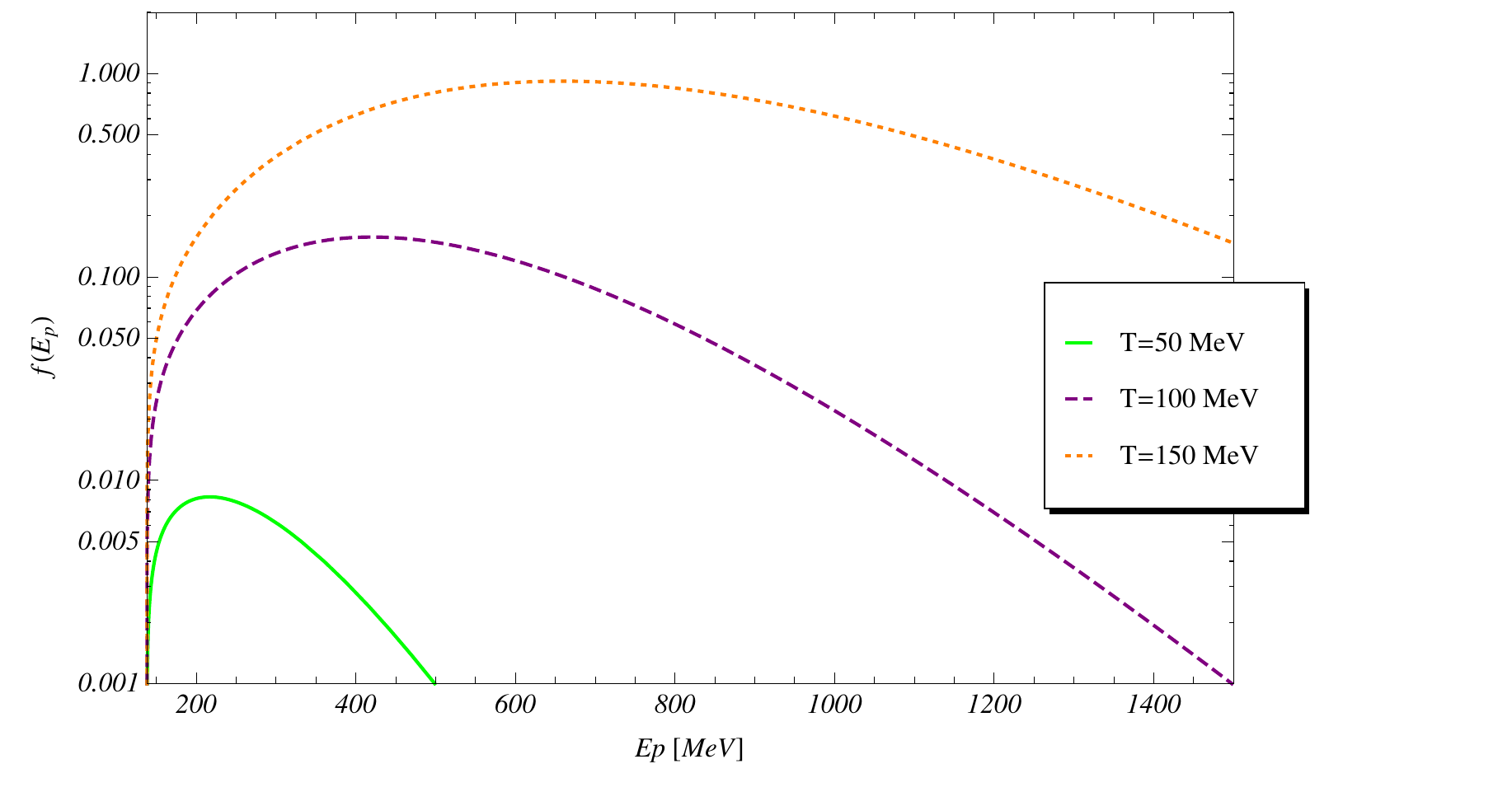}
	\caption[eta pi]{The collisional broadening integrand $f(E_p,T)$~\eqref{integrand} is plotted as a function of the incoming pion energy $E_p$ for three different values of the temperature for NLO large-$N_c\ ChPT$. The collisional broadening is proportional to the area under the graph obtained at the corresponding $T$. This area gets bigger for high temperatures and therefore also the collisional broadening does.}
	\label{fig:integrand}
\end{figure}
Fig.~\ref{fig:integrand} shows the integrand as a function of the pion energy $E_{p}$, for three different values of the temperature --- 50, 100 and 150 MeV. We see that the integrand encompasses higher energies for higher temperatures. Moreover for higher temperatures the collisional broadening integral gets a significant contribution from high pion energies~\cite{Reference2}. This casts doubts on the quantitative reliability of the $ChPT$ results for the collisional width.

\subsection{Resonance Chiral Theory}
\subsubsection{Formalism}
(Large-\textit{$N_{c}$}) $ChPT$ constitutes the appropriate framework to describe the dynamics of the Goldstone bosons at low momenta, but does not include vector and scalar resonances such as the $\rho$, $f_0$, etc. By construction one sticks to energies where the resonances are no excited. These resonances are then integrated out and the respective effects are encoded in the LEC:s of the chiral Lagrangian. It has been shown in~\cite{Ecker:1988te,Donoghue:1988ed} that the LEC:s $L_i$ are dominated by the resonance contributions. Since they are saturated in good approximation by the lowest multiplets of resonances we refer to this effect as ``resonance saturation''.
  
In the combined chiral and large-$N_c$ limit the quasi-Goldstone bosons (including the $\eta'$ meson) become massless. All other meson masses remain finite instead~\cite{Kaiser:2000gs,'tHooft:1974hx,'tHooft:1973jz}.  If we denote with $P$ any quasi-Goldstone boson ($P=\pi,\ K,\ \eta, \ \eta'$) and with $R$ any other meson, then close to the chiral and large-$N_c$ limit we have
$M_P \ll M_R$ 
since $M_R=O(\delta^0)=O(N_c^0)$, i.e.\ resonance masses are not affected by these limits.  
From the relation above it follows that typically the expansion is performed in terms of the dimensionless parameter $\frac{p^2}{M_R^2}$ with $p\sim M_P=O(\delta^{1/2})$. Note that for $M_P=M_{\eta'}$ and $N_c=3$ this expansion parameter is not small at all. There are in fact some resonances lighter than the $\eta'$ meson, namely vector $V(1^{--})$ and scalar $S(0^{++})$ mesons, for which $\frac{M^2_{\eta'}}{M_R^2}>1$~\cite{Agashe:2014kda}. Such resonances must be included if one wants to study the dynamics of the $\eta'$ meson. This is done in $RChT$, where interactions between pseudoscalar mesons are mediated by resonance exchange. Intuitively we can think that when the energy of the process is enough to resolve microscopically a four point vertex of ordinary $ChPT$, then this is stretched out into a propagator. To summarize $ChPT$ calculations become unreliable when the energy available is of the order of the resonance mass.

However, including resonances brings also a disadvantage: the presence of additional degrees of freedom spoils the power counting. This means that there is no longer a systematic EFT. We have to live with a model instead of an EFT but at least we have a well-defined formal low-energy, large-$N_c$ limit as a guideline and anchor. Following the formal power counting rules of large-$N_c\ ChPT$ we still neglect loop diagrams for the calculation of the matrix elements. Essentially we use the previous scattering diagrams and  ``resolve'' low-energy contact interactions.
\subsection*{How to introduce resonances}
In this work we consider the following two reactions: $$\eta' \pi\rightarrow \eta\pi,\qquad \eta' \pi\rightarrow K\bar{K},$$ i.e.\ those contributing to the width increase of the $\eta'$ meson when the latter is placed in a gas of pions. Since we will take into account resonance exchange between the interacting pseudoscalars, we need to include vector and scalar degrees of freedom in the Lagrangian, always preserving chiral symmetry. We introduce the nonet field $S$ which contains the lowest-lying scalar resonances ($J^{PC}=0^{++}$)~\cite{Agashe:2014kda}, see also~\cite{Escribano:2010wt},
\begin{equation*}
S=\begin{pmatrix}\label{S}
\frac{a^0_0}{\sqrt{2}}+\frac{f_0}{\sqrt{2}} & a^+_0 & \kappa^+ \\
a_0^- & -\frac{a^0_0}{\sqrt{2}}+\frac{f_0}{\sqrt{2}} & \kappa^0 \\
\kappa^- & \bar{\kappa}^0 & f_{0s}
\end{pmatrix}
\end{equation*}
and the field $V$ which contains the lowest-lying vector resonances ($J^{PC}=1^{--}$)~\cite{Agashe:2014kda}
\begin{equation*}\label{V}
V_{\mu\nu}=\begin{pmatrix}
\frac{\rho^0}{\sqrt{2}}+\frac{\omega}{\sqrt{2}} & \rho^+ & K^{*+} \\
\rho^- & -\frac{\rho^0}{\sqrt{2}}+\frac{\omega}{\sqrt{2}} & K^{*0} \\
K^{*-} & \bar{K}^{*0} & \phi
\end{pmatrix}_{\mu\nu}.
\end{equation*}
In accordance with the large-$N_c$ spirit we assume that the isoscalars consist of a purely strange and a purely up-down state.
Note that the vector resonances are described by antisymmetric tensor fields~\cite{Gasser:1983yg,Ecker:1988te} instead of vector fields. 

Let us consider once more the large-$N_c \ ChPT$ Lagrangian up to (including) NLO,
\begin{equation}
\mathcal{L}^{ChPT}=\mathcal{L}^{(0)}+\mathcal{L}^{(1)},
\end{equation}
where $\mathcal{L}^{(0)}$ is given by~\eqref{LO} and $\mathcal{L}^{(1)}$ is given by~\eqref{NLO}. The latter consists of terms with $L_i$ and $\Lambda_i$ coefficients, i.e.\
\begin{equation}
\mathcal{L}^{(1)}=\mathcal{L}_{L_{i}'s}+\mathcal{L}_{\Lambda_{i}'s}\,.
\end{equation}
We want to replace $\mathcal{L}_{L_{i}'s}$ with a Lagrangian that contains scalar and vector resonances, namely $\mathcal{L_{\mathrm{res}}}$. Depending on the kind of process that we are studying, different terms of $\mathcal{L_{\mathrm{res}}}$ will contribute since different resonances will be involved. 
Before writing down the expression for the resonance Lagrangian, we need to define the chiral building blocks:
\begin{equation}\label{notation}\begin{split}
&u_\mu= iu^\dagger D_\mu U u^\dagger=u_\mu^\dagger,\\
&\chi_\pm=u^\dagger\chi u^\dagger\pm u\chi^\dagger u,
\end{split}\end{equation}
where $U=u^2=\mathrm{exp}(i\phi)$ as defined in~\eqref{U}. As in the usual notation  $\chi=2B(s+ip)$ and the covariant derivative $D_\mu$ is given in terms of the external sources $a_\mu$ and $v_\mu$~\cite{Ecker:1988te}.

At LO the $RChT$ Lagrangian is the same as the $ChPT$ one~\eqref{LO}, but we can rewrite it using~\eqref{notation}:
\begin{equation}
\mathcal{L}^{(0)}=\frac{F^2}{4}\mathrm{Tr}\left[u_\mu u^\mu+\chi_+\right]-3\tau(\eta_0)^2.
\end{equation}
The NLO part instead differs from $ChPT$: the LEC:s $L_i$ are replaced by resonance exchange. 
In the notation of~\cite{Ecker:1988te} we have:
\begin{equation}\label{Lres}
\mathcal{L_{\mathrm{R}}}=\mathcal{L}_{\mathrm{kin}}(R)+\mathcal{L}_{\mathrm{int}}(R),\quad R=S,V
\end{equation} 
with kinetic and mass terms:
\begin{equation}\label{kinterms}\begin{split}
&\mathcal{L}_{\mathrm{kin}}(S)=\frac{1}{2}\mathrm{Tr}\left[\nabla^\mu S\nabla_\mu S-M^2_S S^2\right],\\
&\mathcal{L}_{\mathrm{kin}}(V)=-\frac{1}{2}\mathrm{Tr}\left[\nabla^\lambda V_{\lambda\mu}\nabla_\nu V^{\nu\mu}-\frac{1}{2} M^2_V V_{\mu\nu}V^{\mu\nu}\right],
\end{split}\end{equation}
where $\nabla_\mu$ is the pertinent chirally covariant derivative for the resonance fields \cite{Ecker:1988te}. We assume that all the particles in the same resonance nonet have the same mass $M_S=M_{a_0}$ for scalar and $M_V=M_{K^{*}}$ for vector resonances. This choice is motivated by the fact that for the scattering processes of interest the only vector meson that is involved is the $K^*$ while among the lowest-lying scalars the $a_0$ meson has the smallest width and therefore the best determined mass~\cite{Agashe:2014kda}. 

The relevant interaction terms are linear in the resonance fields
\begin{equation}\label{SVint}\begin{split}
&\mathcal{L}_{\mathrm{int}}(S)=c_d\mathrm{Tr}\left[Su_\mu u^\mu\right]+c_m\mathrm{Tr}\left[S\chi_+\right],\\
&\mathcal{L}_{\mathrm{int}}(V)=i\frac{G_V}{\sqrt{2}}\mathrm{Tr}\left[V_{\mu\nu}u^\mu u^\nu\right]
\end{split}\end{equation}
and depend on the $RChT$ parameters $c_d$, $c_m$ and $G_V$, whose numerical values will be determined below. 
In principle we could include also axial-vector $A(1^{++})$ and pseudoscalar $P(0^{-+})$ mesons but since they are not exchanged in the reactions that we are  considering, we simply have:
\begin{equation}
\mathcal{L}_\mathrm{res}=\mathcal{L}_S+\mathcal{L}_V.
\end{equation}
Putting together all the terms that build up the Lagrangian used to perform the calculations in the $RChT$ framework we have:
\begin{equation}\label{RChTlag}\begin{split}
\mathcal{L}^{RChT}&=\mathcal{L}^{(0)}+\mathcal{L}_{\mathrm{res}}+\mathcal{L}_{\Lambda_{i}'s}\\
&=\mathcal{L}^{(0)}+\mathcal{L}_S+\mathcal{L}_V+\mathcal{L}_{\Lambda_{i}'s}.
\end{split}\end{equation}

Let us have a closer look at $\mathcal{L}_{\mathrm{res}}$. We expand $U=1+i\phi-\frac{\phi^2}{2}+..$ and keep all the terms that contain up to three fields (i.e.\ two $\phi$'s plus $S$ or $V$). In order to describe QCD we set all the external sources to zero except for $s=\mathcal{M}$ so that we have:
\begin{equation}\begin{split}
\mathcal{L}_S\rightarrow&\,\frac{1}{2}\mathrm{Tr}~[\partial^\mu S\partial_\mu S-M^2_S S^2]\\&+c_d\mathrm{Tr}~[S\partial_\mu\phi \partial^\mu\phi]-\frac{Bc_m}{2}\mathrm{Tr}~\big[S\mathcal{M}\phi^2+2S\phi\mathcal{M}\phi+S\phi^2\mathcal{M}\big]\\&+4Bc_m\mathrm{Tr}~[S\mathcal{M}]
\end{split}\end{equation}
and
\begin{equation}\label{RChTLag}\begin{split}
\mathcal{L}_V\rightarrow&-\frac{1}{2}\mathrm{Tr}~\left[\partial^\lambda V_{\lambda\mu}\partial_\nu V^{\nu\mu}-\frac{1}{2} M^2_V V_{\mu\nu}V^{\mu\nu}\right]\\&+i\frac{G_V}{\sqrt{2}}\mathrm{Tr}~[V_{\mu\nu}\partial^\mu\phi \partial^\nu\phi]\,.
\end{split}\end{equation}
First of all note that the resonance fields are physical fields in the sense that they come already with the correct normalization. Second, but not less important, note that the last term in $\mathcal{L}_S$ is linear in the scalar field $S$. This means that the vacuum expectation value of the field is not zero and therefore $S$ must be shifted~\cite{SanzCillero:2004sk}. As we will see below, this operation gives rise to new interaction terms.
\subsection*{Redefinition of the scalar resonance nonet field}
The linear term in $S$ comes from the interaction term with coupling $c_m$, namely 
$c_m\mathrm{Tr}~(S\chi_+)$.
In fact by expanding $\chi_+$ we have:
\begin{equation*}\label{chi}\begin{split}
\chi_+ = & \,2\chi-\frac{\phi^2\chi+2\phi\chi\phi+\chi\phi^2}{4}\\&+\frac{\phi^4\chi+4\phi^3\chi\phi+6\phi^2\chi\phi^2+4\phi\chi\phi^3+\chi\phi^4}{4!8}+...
\end{split}\end{equation*}
where $\chi=2B\mathcal{M}$.
When this expansion is inserted in $\mathcal{L}_{\mathrm{int}}(S)$ the very first term generates a linear term in $S$, while the others give rise to interaction terms.
This means that the resonance field $S$ has a non-vanishing vacuum expectation value:
\begin{equation*}
\big\langle S\big\rangle_{vev}=\frac{2c_m\chi}{M^2_S}=\frac{4Bc_m\mathcal{M}}{M^2_S}\,.
\end{equation*}
To work around this problem we perform in the scalar field the shift: 
\begin{equation*}
S=\tilde{S}+\big\langle S\big \rangle_{vev}
\end{equation*}
and from now on take into consideration the field $\tilde{S}$ instead, which satisfies
\begin{equation*}
\big\langle \tilde{S}\big \rangle_{vev}=0.
\end{equation*}
When replacing $S$ with $\tilde{S}+\frac{4Bc_m\mathcal{M}}{M^2_S}$ in~\eqref{Lres}, the part of the Lagrangian containing vector resonances is left unchanged. Moreover we get rid of the linear term in $S$ in the scalar part and two new interaction terms pop up. 
If we drop constant terms this is what we have:
\begin{equation}\label{Lstilde}
\begin{split}
\mathcal{L}_{\tilde{S}} =&\,\frac{1}{2}\mathrm{Tr}~\left[\nabla^\mu \tilde{S}\nabla_\mu \tilde{S}-M^2_S \tilde{S}^2\right]+c_d\mathrm{Tr}~\left[\tilde{S}u_\mu u^\mu\right]+c_m\mathrm{Tr}~\left[\tilde{S}\left(\chi_+-2\chi\right)\right]\\
& +\frac{4Bc_mc_d}{M^2_S}\mathrm{Tr}~\left[\mathcal{M}u_\mu u^\mu\right]+\frac{4Bc_m^2}{M^2_S}\mathrm{Tr}~\left[\mathcal{M}\chi_+\right].
\end{split}
\end{equation}
While the terms proportional to $c_m$ and $c_d$ represent interactions of the scalar mesons with the Goldstone bosons (and give rise to three-leg vertices), the terms proportional to $c_mc_d$ and $c_m^2$ are point-interaction and free-field terms. We expand them up to fourth order in $\phi$, to get four-leg vertices.
\subsection*{Resonance saturation}
We illustrate how to obtain the amplitudes for the reactions $\eta' \pi\rightarrow \eta\pi$ and $\eta' \pi\rightarrow K\bar{K}$ starting from the $RChT$ Lagrangian~\eqref{RChTlag}. In the~\ref{B} we show that in the heavy resonance mass limit they are formally identical to the amplitudes obtained in the framework of $ChPT$.
In order to achieve this result we will make use of the relations between the LEC:s and the $RChT$ parameters $c_m,\ c_d$ and $G_V$ derived in~\cite{Ecker:1988te}. 

Every low-energy coupling constant $L_i$ appearing in the NLO $ChPT$ Lagrangian~\eqref{NLO} can be written as a sum
\begin{equation}
L_i(\mu)=\sum_{R=V,S}L_i^R+\hat{L}_i(\mu)
\end{equation}
of resonance contributions\footnote{In principle one should sum also over axial-vector and pseudoscalar resonances but we will ignore them since they do not give relevant contributions to the LEC:s that are of interest for our work.} $L_i^R$ and a scale-dependent remainder $\hat{L}_i(\mu)$. Resonance saturation assumes that $\hat{L}_i$ is solely caused by loops and therefore suppressed in the $1/N_c$ expansion --- $L_i^R\sim N_c,\ \hat{L_i}\sim N_c^0$. Since the evaluation of loops depends on the renormalization scale $\mu$, the values for $\hat{L}_i(\mu)$ must be chosen such that observables are $\mu$ independent. Below we will compare our values for the $L_i$'s to other works. To do so we have to choose a renormalization point since the $L_i$'s are not observables. We expect eventually to observe the resonance dominance when $\mu$ is close to the resonance region and therefore one usually assumes $\hat{L}_i(\mu)\ll L_i^R$ for $\mu\approx M_\rho=775$ MeV~\cite{Ecker:1988te}.

Among all the relations derived in~\cite{Ecker:1988te}, the ones that we need are:
\begin{equation}\label{LECrel}\begin{aligned}
L_2^V&=\frac{G^2_V}{4M^2_V}, & L_3^V&=-3L_2^V, & L_5^V&=0, & L_8^V&=0, \\
L_2^S&=0, & L^S_3&=\frac{c_d^2}{2M^2_S}, & L_5^S&=\frac{c_dc_m}{M^2_S}, & L^S_8&=\frac{c_m^2}{2M^2_S}.
\end{aligned}\end{equation}
The first row regards the vector contributions to the LEC:s while in the second row there are the scalar ones. 
The saturation of the LEC:s implies that basically there is ``no room'' for contributions from anything else than meson resonances, i.e.\ $\hat{L}_i(M_\rho)\simeq 0$.

\subsection*{Determination of the $RChT$ parameters $c_d,\ c_m$ and $G_V$}
In the framework of $RChT$ we use a Lagrangian where in the spirit of resonance saturation~\cite{Ecker:1988te} the LEC:s have been replaced by resonance exchange processes. The $RChT$ Lagrangian is however given in terms of other parameters: $c_m,c_d$ and $G_V$. Using the mass of the $a_0$ resonance, it is straightforward to determine the value of $c_m$ and $c_d$ given our previous determination of the LEC:s. In fact, once we know the numerical values of $L_5$ and $L_8$, inverting the relations~\eqref{LECrel} we obtain:
\begin{equation}
c_m=\sqrt{2M^2_SL_8}=50.3 \ \text{MeV} \qquad\text{and}\qquad c_d=\sqrt{\frac{M^2_SL_5}{c_m}}=41.9\text{ MeV}.
\end{equation}
One way to find the value of $G_V$ is to consider the decay $K^*\rightarrow K\pi$. For simplicity we look at the positively charged strange vector meson.
As a consequence of isospin symmetry, we find:
\begin{equation*}\begin{split}
\Gamma_{K^{*}\rightarrow K \pi}&=\Gamma_{K^{*+}\rightarrow K^+ \pi^0}+\Gamma_{K^{*+}\rightarrow K^0 \pi^+} \\&=3\Gamma_{K^{*+}\rightarrow K^+ \pi^0}
\end{split}\end{equation*}
where:
\begin{equation}
\Gamma_{K^{*+}\rightarrow K^+ \pi^0}=\frac{G^2_V[(M^2_{K^*}-(M_\pi-M_K)^2)(M^2_{K^*}-(M_\pi+M_K)^2)]^{3/2}}{192\pi F_K^2F_\pi^2 M^3_{K^*}}.
\end{equation}
The experimental value of the partial decay width $\Gamma_{K^{*}\rightarrow K \pi}$~\cite{Agashe:2014kda} is:
\begin{equation*}
\Gamma_{K^{*}\rightarrow K \pi}\simeq 47 \text{~MeV}
\end{equation*}
from which we can derive 
\begin{equation}
G_V=70.7 \text{~MeV}.
\end{equation} 

\subsection*{The remaining LEC:s --- $L_2$ and $L_3$}
At this point we can finally determine the numerical values of the last two LEC:s. We know that they are related to the $RChT$ parameters by the relations~\eqref{LECrel} which lead to
\begin{equation}
L_2=\frac{G_V^2}{4M^2_V},\qquad L_3=\frac{c_d^2}{2M^2_S} -\frac{3G_V^2}{4M^2_V}
\end{equation}
where $M_V$ indicates the mass of the vector resonance ($M_{K^*}$ in our case) and $M_S$ stands for the mass of the scalar resonance ($M_{a_0}$ in our case).
For $c_d,c_m$ and $G_V$ we use the values just determined previously. Then we get:
\begin{equation}
L_2=1.56\cdot10^{-3} \qquad\text{and}\qquad L_3=-3.76\cdot10^{-3}.
\end{equation}
Once we know all the values of the LEC:s and coupling constants $c_d,c_m$ and $G_V$, we have predictive power.

We compare in Table~\ref{tab:compare} the values for the LEC:s in the large-$N_c$ limit derived in this work with the values used in~\cite{Reference2} and with those presented in~\cite{Pich:2008xj,Bijnens:2014lea}.
\begin{table}[h]
\centering
\begin{tabular}{l r r r r r}
\toprule \hline
LEC:s    &   & Ref.~\cite{Reference2}  & Ref.~\cite{Bijnens:2014lea} & Ref.~\cite{Pich:2008xj} \\
\midrule
$L_2$       & 1.56     & 1.8       &1.6(2)    &1.8 \\
$L_3$       & $-3.76$  & $-4.31$   &$-3.8(3)$ &$-4.3$\\
$L_5$       & 2.19     & 2.25      &1.2(1)    &2.1\\
$L_8$       & 1.31     & 1.03      &0.5(2)    &0.8\\
\bottomrule
\end{tabular}
\caption{The values of the LEC:s (in units of $10^{-3}$) from various references are compared with those determined in this work (second column).}
\label{tab:compare}
\end{table}
A part from the $L_8$ value, there seems to be a quite good agreement between the results obtained with different strategies. Luckily, the LEC:s that play the most important role in our calculations are $L_2$ and $L_3$ (see $ChPT$ scattering amplitudes ~\eqref{etapi},~\eqref{kaons}). These are very close to the values recently presented in~\cite{Bijnens:2014lea}.
An estimate of the errors  would require a calculation beyond NLO, which is beyond the scope of the present work.
\subsubsection{Results}
\begin{figure}[h]
\centering
   \includegraphics[width=12cm]{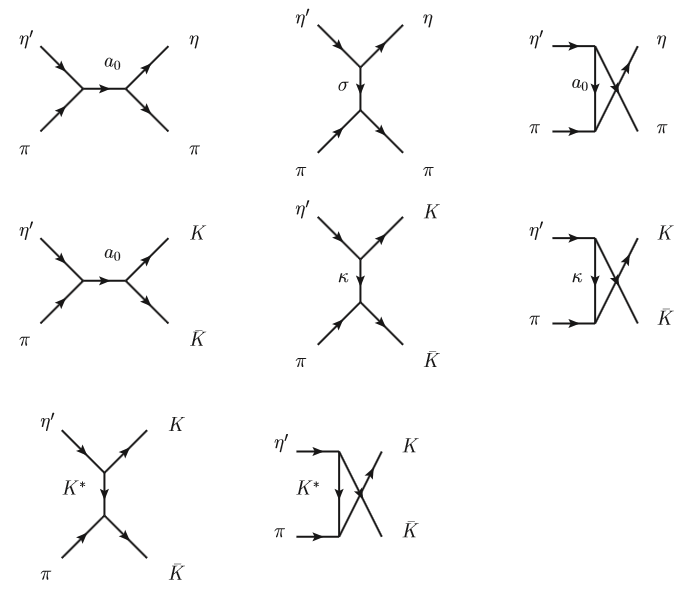}
   \caption{Additional $RChT$ diagrams contributing to the in-medium width of the $\eta^\prime$ .}\label{RChT}
 \end{figure}

The scattering amplitudes for the reactions $\eta'\pi\rightarrow\eta\pi$ and $\eta'\pi\rightarrow K\bar{K}$ derived from the $RChT$ Lagrangian can be found in \ref{B}. The corresponding diagrams are those already obtained in $ChPT$, i.e.\ the four-point vertex diagrams of Fig.~\ref{ChPT}, together with those shown in Fig.~\ref{RChT} where resonances are exchanged. As a consequence of introducing resonances, we find in the scattering amplitude the corresponding propagators. For a resonance of mass $M$ and momentum $k$ this is proportional to $1/(k^2-M^2)$. It follows that the energy dependence of the matrix element will be suppressed, pushing down the values of the cross sections in particular for high energies. This would then decrease the total width addition at high temperatures, leading to a longer lifetime of the $\eta'$ with respect to the $ChPT$ case.
Regarding the $\eta'\pi\rightarrow\eta\pi$ process, due to the quantum numbers of the pseudoscalar mesons involved, the only resonances that can be exchanged are scalar: the $a_{0}(980)$, a light $f_{0}$ (or $\sigma$) and a strange $f_{0s}$, even though the latter does not contribute. We refrain from assigning a physical state to the $f_0$. In the large-$N_c$ spirit we will use the same mass for $a_0$ and $f_0$. We will come back to this point when discussing our results. Regarding the $\eta'\pi\rightarrow K\bar{K}$ process, both scalar and vector resonances can be exchanged. In particular, the $a_{0}(980)$ and the $K^*_0(800)$ (or $\kappa$) among scalar mesons and the $K^*(892)$ among vector mesons. For more details on the derivation of the vertices see~\cite{myThesis}. 

\begin{figure}
	\centering
		\includegraphics[width=10cm]{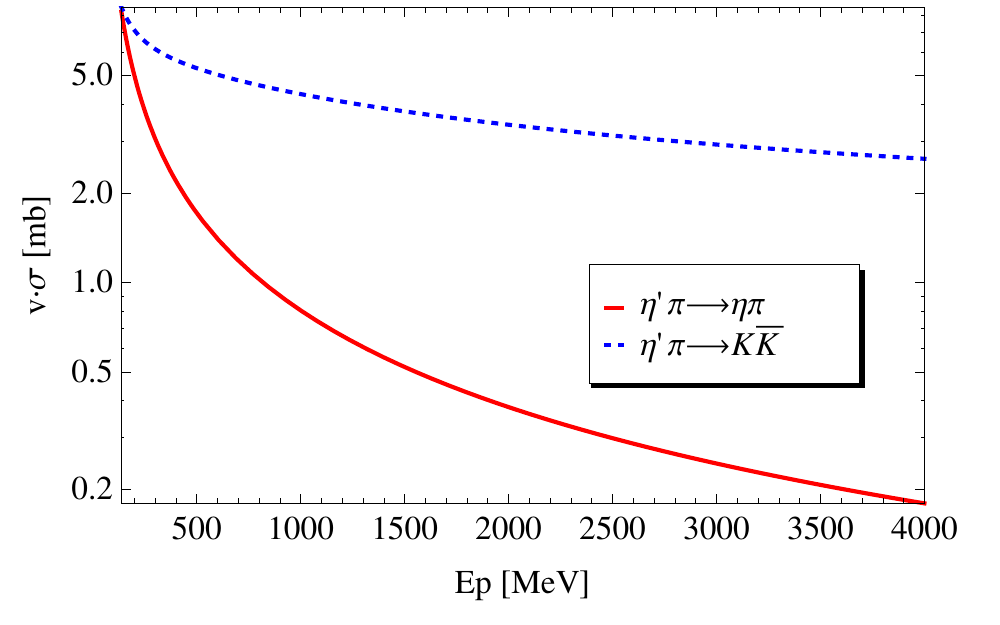}
	\caption{The reaction rates $v\cdot\sigma$ for the processes $\eta' \pi\rightarrow\eta\pi$ and $\eta' \pi\rightarrow\bar{K}K$ at NLO in $RChT$, plotted as functions of the incoming pion energy $E_p\ge M_\pi$ in the rest frame of the $\eta'$. The range of values for $E_p$ is wider than that used to plot the $ChPT$ reaction rates in Fig.~\ref{rateChPT} so that it is immediate to see that the $RChT$ cross section for $\eta'\pi\rightarrow\eta\pi$ goes to zero for high energies.}
	\label{rateRChT}
\end{figure}

The reaction rates of interest for the calculation of the collisional broadening are shown as a function of $E_p$ in Fig.~\ref{rateRChT}. We note big differences with respect to the $ChPT$ results of Fig.~\ref{rateChPT}. First of all the reaction rates decrease as the pion energy increases. In particular the reaction rate for the process $\eta'\pi\rightarrow\eta\pi$ decreases faster than that with two kaons in the final state. The latter keeps decreasing until it reaches a minimum and then starts growing again but very slowly~\cite{myThesis} (not visible within the energy range of Fig.~\ref{rateRChT}).
\begin{figure}
	\centering
		\includegraphics[width=11cm]{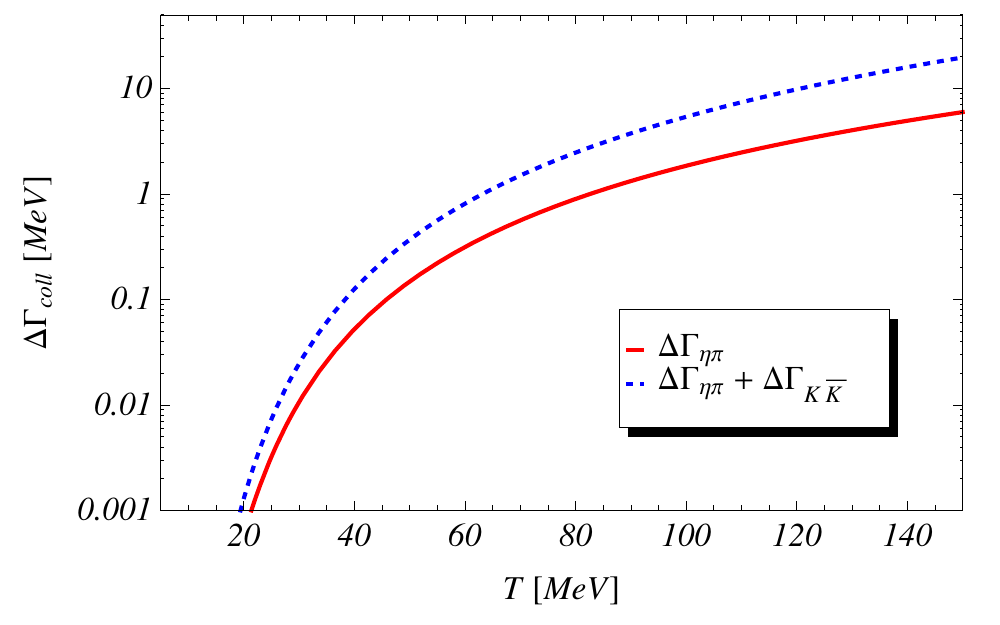}
	\caption{Addition to the $\eta'$ width as a function of the temperature at NLO in $RChT$.}
	\label{fig:GammaRChT}
\end{figure}

This means that in the $RChT$ case it is really important to consider the $\bar{K}K$ final state reactions because their contribution to the collisional broadening is dominant. Finally note that the reaction rates do not diverge at $E_p=M_\pi$ (as it could seem from Fig.~\ref{rateRChT}). What we see is indeed the tail of the subthreshold $a_0$-resonance. In fact the increase with lower energies is due to the presence of the pole in the $a_0$ propagator (s-channel) at $M_{a_0}=980$ MeV which is below but close to threshold $M_{\eta'}+M_\pi=1096$ MeV.

In the framework of $RChT$ we obtain the matrix elements  $\mathcal{M}_{\eta' \pi\rightarrow \eta\pi}^{RChT}$ and $\mathcal{M}_{\eta' \pi\rightarrow \bar{K}K}^{RChT}$ (see~\eqref{RChTetapi},~\eqref{kaonRChT}). The latter has an energy dependence while the former is, at least for large energies, constant at NLO. To understand why, let us look at~\eqref{RChTetapi}. For large $\textit{s, t}$ and $\textit{u}$, the dominant terms are:
\begin{equation*}
\frac{c_{d}^{2} s^{2}}{(M^{2}-s)} + \frac{c_{d}^{2} t^{2}}{(M^{2}-t)} + \frac{c_{d}^{2} u^{2}}{(M^{2}-u)} \xrightarrow{s,t,u \gg M^{2}} -c_{d}^{2} (s+t+u)
\end{equation*}
Since the combination $\textit{s} + \textit{t} + \textit{u}$  is equal to the sum of the squared masses of the four mesons involved in the process, it is  indeed a constant. This explains why the matrix element becomes constant for large energies in $RChT$. It follows that the corresponding reaction rate decreases for higher pion energies, as shown in Fig.~\ref{rateRChT}. 
This ``nice'' behavior for high pion energies is unexpected, especially because it does not indicate any break down of the low-energy theory. It can be expected that $RChT$ has a better high-energy behavior than $ChPT$, because constants are replaced by propagators. But naively one would expect that an NLO matrix element changes from $\sim s^2$ to $\sim s$, not to a constant at large $s$. 
To summarize: for the processes studied here $RChT$ has the low-energy limit of an EFT, but a much better high-energy behavior.
 
Fig.~\ref{fig:GammaRChT} shows the main result of this paper --- the increase of the $\eta'$ width at NLO of $RChT$. For a temperature $T\approx 120$ MeV we find a width increase of $\Delta\Gamma_{\mathrm{coll}}\approx10$ MeV, smaller than in the $ChPT$ case, but still comparable with the inverse lifetime of the fireball created in a heavy-ion collision. 
\begin{figure}[h]
	\centering
		\includegraphics[width=13cm]{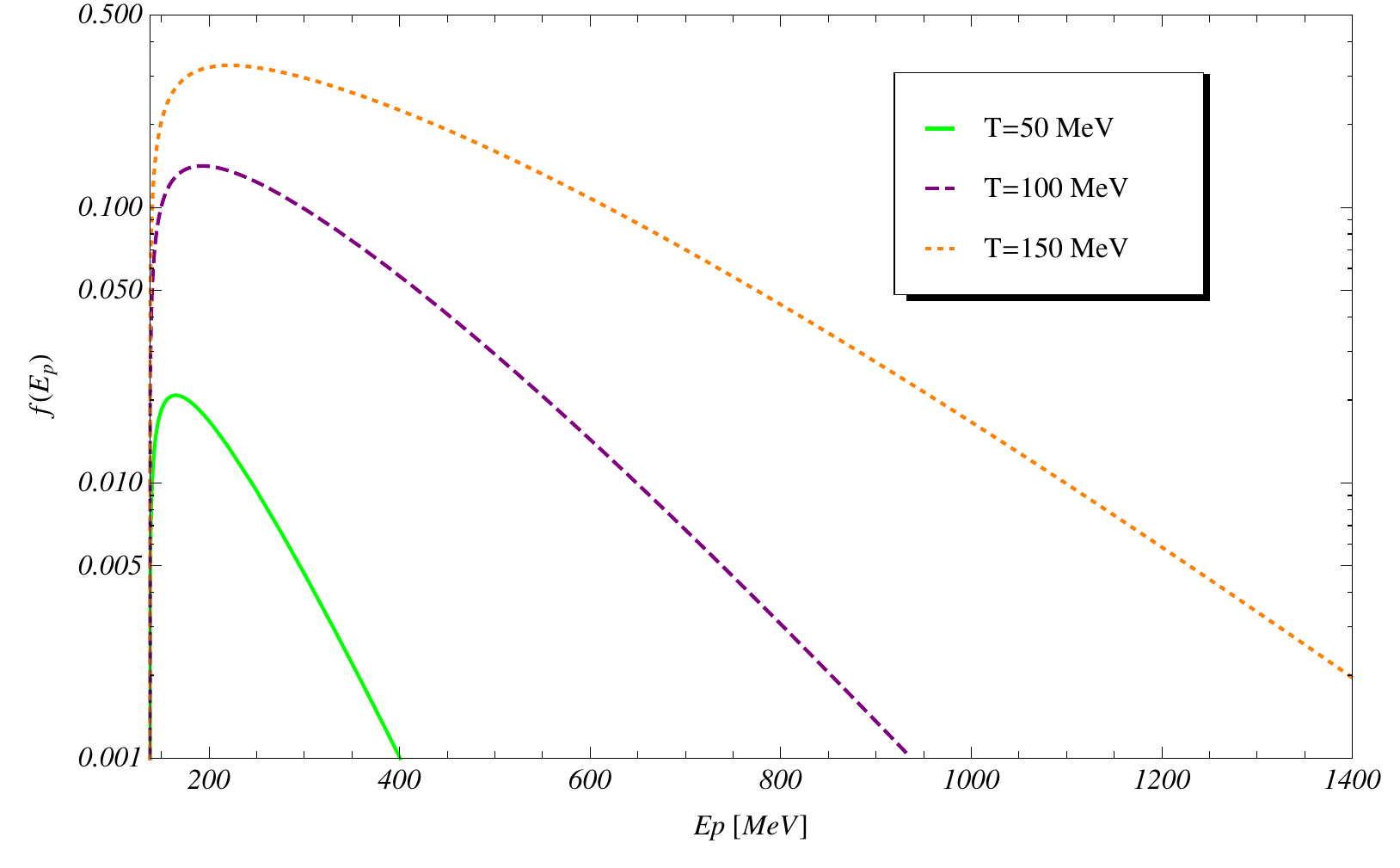}
	\caption[eta pi]{The collisional broadening integrand $f(E_p,T)$~\eqref{integrand} is plotted as a function of the incoming pion energy $E_p$ for three different values of the temperature for NLO $RChT$. Compared to the $ChPT$ case (Fig.~\ref{fig:integrand}) the integrand decreases very fast down to zero for high $E_p$ and therefore the collisional broadening does not get a big contribution from the high-$E_p$ region.}
	\label{fig:integrandRChT}
\end{figure}

We plot in Fig.~\ref{fig:integrandRChT} the integrand from the collisional broadening integral as a function of the pion energy $E_{p}$, for three different values of the temperature --- 50, 100 and 150 MeV.
We see that the contribution to the collisional broadening integral decreases very fast for high pion energies as a consequence of the behavior of the $RChT$ cross sections. Thus the dominant contributions to the collisional width come from low energies where $RChT$ should produce quantitatively reliable results. 

\subsection*{Scalar-meson masses and $RChT$ amplitudes}
\begin{figure}[ht]
	\centering
		\includegraphics[width=13cm]{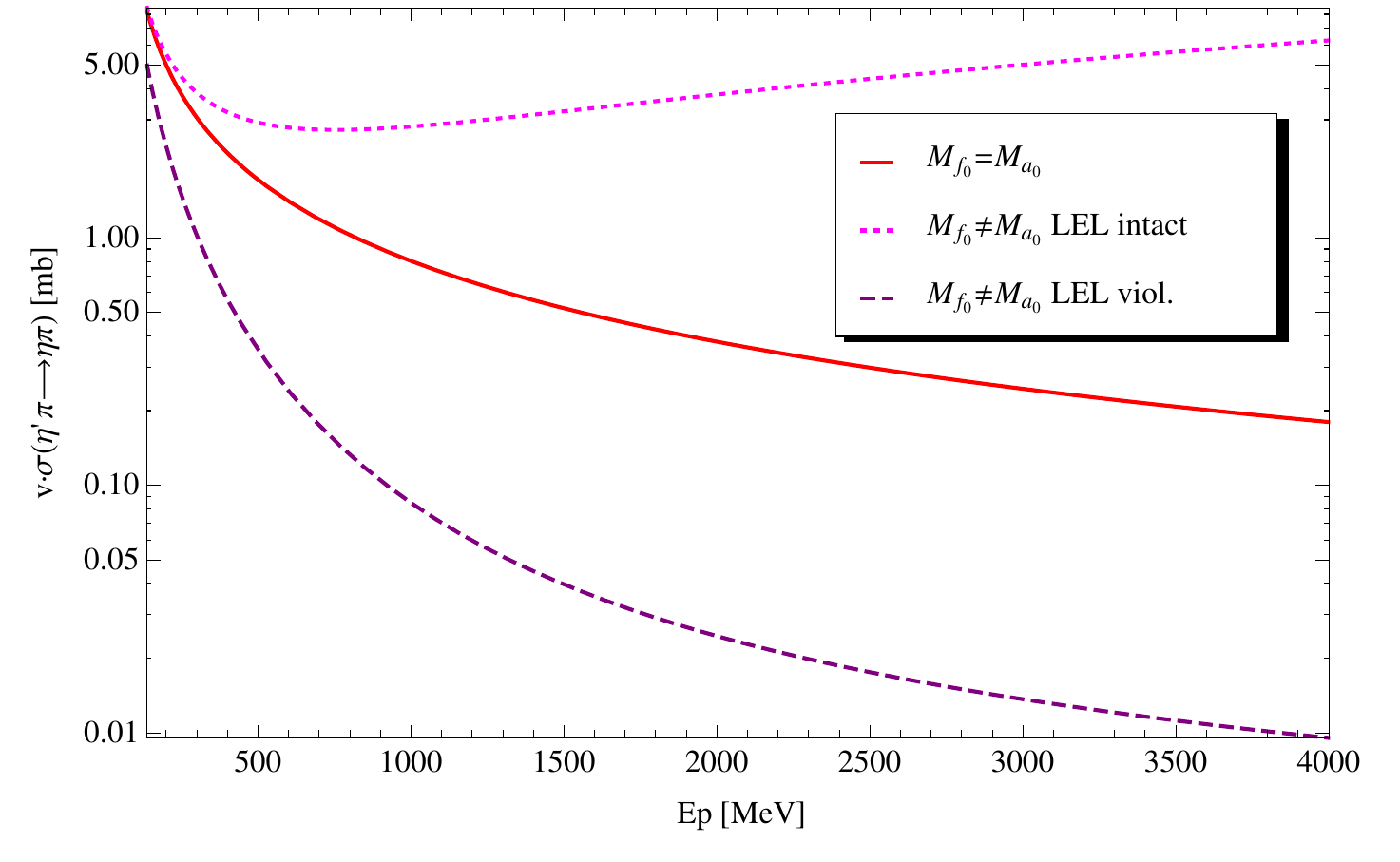}
	\caption{The reaction rate for the process $\eta' \pi\rightarrow\eta\pi$ goes to zero faster if we use for the $f_0$ resonance the real value of its mass and not $M_{a_0}$ (dashed line). This modification violates the low-energy limit and therefore is referred to as LEL viol. If we modify the propagator of the $f_0$ resonance as in~\eqref{prop} (LEL intact), the reaction rate does not vanish but increases slowly for high energies (dotted line).}
	\label{MassEtapi}
\end{figure}

In the following we discuss some variations which in part will change the high-energy behavior. So far we have chosen to put all the scalar resonance masses equal to the mass of the $a_0$ resonance. This choice is justified by the fact that mass splitting effects in this nonet are suppressed by the quark masses or $1/N_c$, i.e.\ these are effects beyond our NLO calculations. But having the same masses is not entirely realistic. Since we know which resonance is exchanged in each channel, we can replace $M_{a_{0}}$ with $M_{f_0}$ or $M_\kappa$ in the propagator and see what changes. The first replacement assumes that the $f_0(500)$~\cite{Agashe:2014kda} is a quark--antiquark state. We postpone this discussion to the summary in Section~\ref{Sec4}. Here we want to see first the impact of changing the masses of the scalar resonances. Note that by doing this the low-energy limit is violated (LEL viol.). When applying this replacement to the $\eta'\pi\rightarrow\eta\pi$ amplitude, we find that the corresponding reaction rate still goes to zero for high pion energies, decreasing indeed faster than before. This is shown in Fig.~\ref{MassEtapi} (dashed line).
On the other hand we can also replace the mass but keep the low-energy limit the same (LEL intact):
\begin{equation}\label{prop}
\frac{1}{t-M^2_{a_0}}\rightarrow \frac{M_{f_0}^2}{M_{a_0}^2\cdot(t - M_{f_0}^2)}.
\end{equation} 
This time we obtain a reaction rate that after an initial decrease starts growing again, but extremely slowly. This does not constitute a problem since at very high energies the reaction rate is still small as can be seen in Fig.~\ref{MassEtapi} (dotted line). 

\begin{figure}[ht]
	\centering
		\includegraphics[width=13cm]{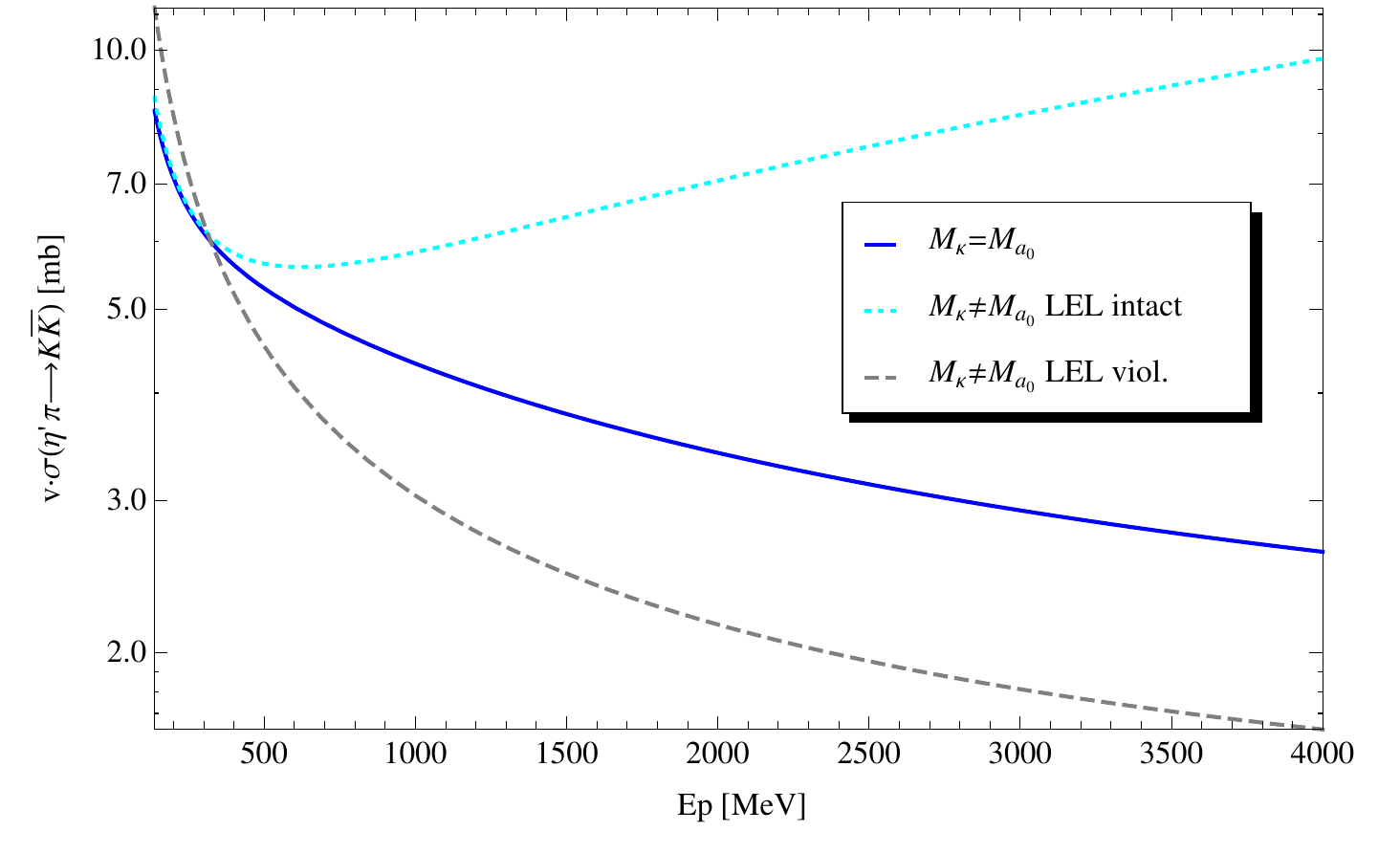}
	\caption{The reaction rate for the process $\eta' \pi\rightarrow\bar{K}K$ decreases faster if we use for the $\kappa$ resonance the real value of its mass and not $M_{a_0}$ (dashed line). This change spoils the low-energy limit (LEL viol.). If we modify the propagator of the $\kappa$ resonance as in~\eqref{prop1} (LEL intact), the reaction rate increases faster instead (dotted line). Even if not visible within this energy range, in both cases the reaction rates do not vanish for high energies as it is also in standard $RChT$.}
	\label{massK}
\end{figure}

We can then apply the same reasoning to $\mathcal{M}_{\eta' \pi\rightarrow \bar{K}K}^{RChT}$. As before we can modify the matrix element by replacing the scalar mass $M_{a_0}$ with $M_\kappa$ in the propagator for the t- and u-channel (LEL viol.) since the $\kappa$ meson is exchanged in those channels. Alternatively we can modify the propagator such that the low-energy limit remains constant (LEL intact):
\begin{equation}\label{prop1}
\frac{1}{t,u-M^2_{a_{0}}}\rightarrow \frac{M^2_\kappa}{M^2_{a_{0}}(t,u-M^2_\kappa)}.
\end{equation}
The results obtained resemble those described before for $\mathcal{M}_{\eta' \pi\rightarrow \eta \pi}^{RChT}$ as we can see in Fig.~\ref{massK}. The main difference is that in this case already the standard $RChT$ amplitude does not go to zero for high energies and the same holds for all its modified versions. The reaction rates start growing again at very high energies; however this cannot be seen in Fig.~\ref{massK} since it occurs outside the plotted energy range.

\begin{figure}[h]
	\centering
		\includegraphics[width=0.9\textwidth]{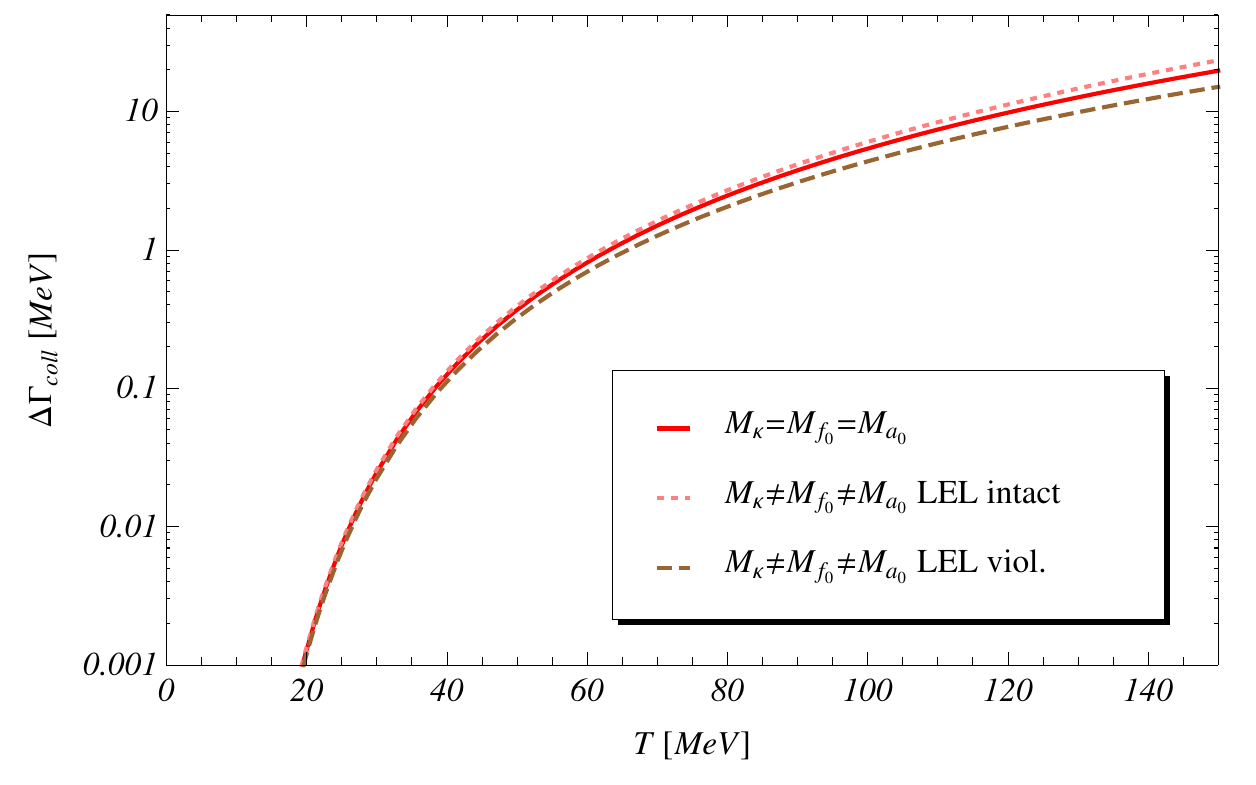}
	\caption{Comparison of the total collisional width from $\eta'\pi\rightarrow\bar{K}K$ and $\eta'\pi\rightarrow\eta\pi$ reactions in NLO $RChT$ obtained using only one value for the scalar resonance masses (full line) and using different values for the masses of the three resonances involved -- $a_0,\ f_0$ and $\kappa$ -- keeping the LEL intact (dotted line) or not (dashed line).}
	\label{GMas}
\end{figure}

It is interesting to plot the total width increase due to $\eta' \pi\rightarrow\eta\pi$ and $\eta'\pi\rightarrow\bar{K}K$ collisions obtained using the modified matrix elements which contain all the different values of the resonance masses ($M_{a_0},\ M_{f_0}, \ M_\kappa$). In Fig.~\ref{GMas} we compare the new results with the standard $RChT$ prediction. The dotted line refers to the case where the propagators have been modified according to~\eqref{prop} and~\eqref{prop1} (LEL intact), while the dashed line corresponds to the other procedure (LEL viol.). As we can see they do not differ much from the case where all the masses are considered to be equal to $M_{a_0}$. This supports the choice of keeping only one mass for all the resonances in the same nonet and demonstrates the robustness of the results. We can briefly comment on Fig.~\ref{GMas} recalling that it is obtained from the reaction rates plotted in Fig.~\ref{MassEtapi} and Fig.~\ref{massK}. The mass modifications that violate the LEL produce a slightly smaller width increase, in agreement with the fact that the corresponding reaction rates are also smaller than the standard $RChT$ ones. On the other hand the mass modifications that leave the LEL intact give a slightly larger width increase due to the fact that the corresponding reaction rates grow for high energies.

\begin{figure}[h]
	\centering
		\includegraphics[width=0.9\textwidth]{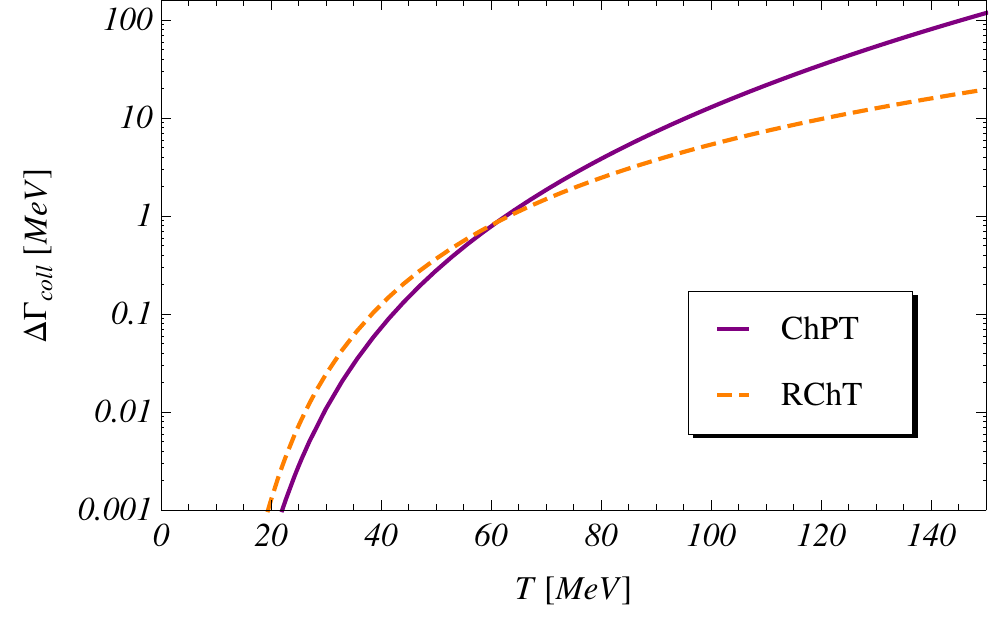}
	\caption{Comparison of the addition to the $\eta'$ width from $\eta'\pi\rightarrow\bar{K}K$ and $\eta'\pi\rightarrow\eta\pi$ reactions in NLO $ChPT$ and $RChT$.}
	\label{compGamma}
\end{figure}

Finally we compare the total width increase of the $\eta'$ meson in $ChPT$ and $RChT$. Fig.~\ref{compGamma} shows that for high temperatures there is almost an order of magnitude of difference between the two. 
The NLO results in $RChT$ differ from those in $ChPT$ in many aspects mainly due to the fact that the $RChT$ cross sections drop for large pion energies, instead of growing. Based on Fig.~\ref{compGamma} we conclude that there is no range where $RChT$ and $ChPT$ results agree, not even at  very low temperatures. According to our previous considerations, $RChT$ is more reliable than $ChPT$ and therefore we draw our conclusions relying on the $RChT$ results. 

We find that already for temperatures around $T\approx 120$ MeV the width increase corresponds to $\Delta\Gamma_{\mathrm{coll}}\approx10$ MeV.
 Even for temperatures around $T_c\approx150$ MeV the width increase is $\Delta\Gamma_{\mathrm{coll}}\approx20$ MeV, still comparable with the inverse lifetime of a fireball. However $RChT$ is not the appropriate tool to explore this temperature region. In fact, close to the phase transition, the approximation of the medium with a pion gas becomes unreliable and therefore we cannot trust our results anymore. Instead, we can draw meaningful conclusions as long as we refer to lower temperatures: we find for $T<T_c$ a considerable width increase, which implies a shorter lifetime for the in-medium $\eta'$ meson. This result suggests that it might be possible to study the $\eta'$ in heavy-ion collision experiments.

\section{Summary and Discussion}
\label{Sec4}
Our interest in the $\eta'$ meson is motivated by the fact that its properties are largely influenced by the $U(1)_A$ anomaly of QCD.
In this paper we have calculated the width increase of the $\eta'$ meson due to interactions with a  gas of pions as a function of the temperature.

Based on $ChPT$ and $RChT$ calculations, we presented predictions for the thermal changes of the $\eta'$ width. Even if the most drastic changes are expected close to the phase transition ($T_c\approx 170$ MeV), one does not have solid tools to perform the calculations at these temperatures. With $ChPT$ (or extension thereof) one cannot address the region around the phase transition, but one can look for the onset of changes at low temperatures. The advantage is that an EFT offers a systematic approach and therefore produces reliable results. Large-$N_c \ ChPT$ permits the formal inclusion of the $\eta'$ meson, which becomes massless in the combined large-$N_c$ and chiral limit. Knowing that large-$N_c \ ChPT$ is formally systematic at large enough $N_c$, an immediate question arises: is $N_c=3$ large enough? There are indications that the large-$N_c$ expansion still makes sense~\cite{Witten:1980sp,DiVecchia:1980ve,Kaiser:2000gs,'tHooft:1973jz,'tHooft:1974hx,Leutwyler:1997yr} if one can handle the obvious flaws (e.g. $M_{\eta'}>M_{R}$). This is what $RChT$ tries to do.

We have considered the two processes that are relevant to the collisional broadening: $\eta'\pi\rightarrow\eta\pi$ and $\eta'\pi\rightarrow\bar{K}K$. The calculations have been performed first in the framework of large-$N_c \ ChPT$~\cite{Reference2}.  To calculate cross sections and first of all scattering amplitudes we have used the NLO large-$N_c\ ChPT$ Lagrangian~\eqref{NLO}. In this framework one has to deal only with point-interaction diagrams which lead to unreliably large scattering cross sections for the energy range of interest. 

Subsequently we have repeated the calculations in the framework of $RChT$ \cite{myThesis}, using an NLO Lagrangian that includes explicitly the effect of resonance exchange in place of the LEC:s. We have included the scalar and vector resonances through nonet fields.  When considering the reactions mentioned above, one obtains also exchange diagrams in addition to point-interaction diagrams. Fortunately the obtained input for the collisional-width formula is not very sensitive to high energies where the framework would become unreliable. The $RChT$ results indicate still a sizeable increase of the width yet smaller than in the $ChPT$ case. At a temperature $T\approx 120$ MeV we have $\Delta\Gamma_{\mathrm{coll}}\approx 10$ MeV, still comparable with the inverse lifetime of a fireball produced in heavy-ion collisions. This information can be useful in the prospect of performing spectroscopy of the $\eta'$ in heavy-ion collisions.

Finally, some discussion is in order concerning the significance of scalar quark-antiquark states in the 1 GeV mass range:
In most of our calculations including resonances we have used the observed $a_0(980)$ meson \cite{Agashe:2014kda} to set the mass of the 
lowest-lying large-$N_c$ scalar meson nonet; see also \cite{Ecker:1988te}. On the other hand, since decades it is discussed in the literature whether the observed 
scalar states with masses below and around 1 GeV are actually dominantly quark-antiquark states or contain a significant amount 
of non-minimal quark content (compact tetraquarks, diquark--anti-diquark systems, hadronic molecules) or of a glueball; see, e.g., \cite{Baru:2003qq,Nebreda:2011cp,Nieves:2011gb,Danilkin:2012ua} and references therein. 
In fact, (pure) non--quark-antiquark states would not contribute in the way as we have used them to the 
saturation of LEC:s of large-$N_c$ $ChPT$. On the other hand, we have not used detailed 
information about the scalar resonances besides the nonet mass. The coupling constants have been determined by matching to large-$N_c$ $ChPT$. In addition, by varying the masses of the scalars while keeping the low-energy limit constant, we have checked that our results are not very sensitive to such variations. The only thing which matters for our framework is the question if 
there is a scalar nonet around 1 GeV in the limit of a large number of colors. There is no first-principle answer from 
QCD yet, but there are indications from (large-$N_c$) lattice QCD \cite{Bali:2013kia} for such a nonet. In addition, combining 
high-energy constraints with algebraically realized chiral symmetry \cite{Weinberg:1969hw} suggests a degeneracy of scalar and vector 
mesons, $M_S = M_V \approx M_\rho \approx 1 \,$GeV. At present we regard this as enough indications for the existence of 
a low-lying scalar nonet to justify our approach. 
We note, however, that the possible existence/importance of non--quark-antiquark states in the 1 GeV range or below in our world 
of three colors can provide a source of problems for any large-$N_c$ framework that claims significance for the real world. 
For instance, this applies also to ordinary (not large-$N_c$) $ChPT$ if the size of LEC:s is 
estimated from large-$N_c$ resonance saturation; see \cite{Bijnens:2014lea} and references therein. If only scalar states from higher 
up in mass are considered for the LEC:s then the influence of the physical low-lying scalar resonances is 
relegated to the loop diagrams where a perturbative treatment might not be enough to account for these resonant effects. 
Possible future improvements concerning our scattering amplitudes will be discussed in the outlook. 

\section{Outlook}
\label{Sec5}
Concerning the in-medium properties of the $\eta'$ meson, there are still many interesting aspects to be studied. In the present work we have investigated one issue: the lifetime. The ambition of the follow-up works is to determine for the $\eta'$ meson the respective thermal changes of mass and coupling strength to the axial-vector current\footnote{The latter two issues involve also the change in the mixing angles between the $\eta$ and the $\eta'$ meson.} in the $RChT$ framework. 
There is an intriguing argument suggesting 
a drastic change of the mass of the $\eta'$ meson \cite{Jido:2011pq}. We shall repeat this line of reasoning 
in the following and stress that it is solely chiral restoration that enters. 
One does not need any argument related to the chiral anomaly; see also \cite{Cohen:1996ng,Lee:1996zy}. 
Let's consider the chiral limit again, i.e.\ set 
the three lightest quark masses to zero. When the spontaneous breaking of $SU_A(3)$ is restored in a medium, one can order 
the excitations according to the product group $SU_L(3) \times SU_R(3)$. For instance, the octets with vector and axial-vector 
quantum numbers, which form separate multiplets in the vacuum, fuse to a 16-plet: If we characterize states by their 
multiplicities with respect to $SU_L(3)$ (first number) and $SU_R(3)$ (second number), then vector and axial-vector mesons 
are obtained by superpositions of $(3,1) \times (\bar 3,1) = (8,1) + (1,1)$ and $(1,3) \times (1,\bar 3) = (1,8) + (1,1)$. 
Since parity is conserved by the strong interaction the states of $ (8,1)$ and $(1,8)$ are degenerate. They build the 16-plet 
of vectors and axial-vectors mentioned above. 

What happens to the Goldstone boson octet? Naively one might assume that also here two octets (pseudoscalar and scalar) form 
a 16-plet, like for the (axial-)vector case. However, one find an 18-plet instead, since in the (pseudo-)scalar case the 
left- and right-handed parts of the quark fields are intertwined: $(3,1) \times (1,\bar 3) = (3, \bar 3)$ and 
$(\bar 3,1) \times (1,3) = (\bar 3, 3)$. Again, parity links the two representations leading to one degenerate 18-plet. 
Thus, after chiral restoration the 8 states with the quantum numbers of the Goldstone bosons are part of a multiplet 
that contains --- besides scalars --- also the flavor-singlet pseudoscalar state. In short, in the chiral limit the $\eta'$ meson becomes degenerate with the Goldstone bosons at the point of chiral restoration. 
Of course, after chiral restoration took place there are no Goldstone bosons, i.e.\ massless states, any more. 
However, we expect precursor 
effects to happen already on the way towards full restoration. In this temperature regime below chiral restoration the masses 
of the Goldstone bosons are protected. This suggests that the mass of the $\eta'$ meson is dragged down towards the pseudoscalar
octet. Of course, away from the chiral limit this argument should be regarded as a qualitative one. By no means the mass of 
the $\eta'$ meson will vanish. But it might be strongly reduced as a function of temperature. It will be interesting to see if there are precursor effects at low temperatures.

Regarding the contents of our present work, we also have some suggestions for further developments. In principle our formalism to determine the in-medium width of the $\eta'$ meson at low temperatures can be decomposed in 
two parts: The low-density approximation that relates the width to the $\eta'$-$\pi$ scattering amplitude and the actual 
determination of this amplitude using large-$N_c$ considerations with or without explicit resonances. Of course, the biggest 
uncertainties rest in the latter part. We have discussed the problems appearing, on the one hand, when using pure large-$N_c$ 
$ChPT$ in an energy range where resonances constitute active degrees of freedom, and on the other hand, the 
model dependence (absence of power counting) emerging from the inclusion of these resonance degrees of freedom. 
In the absence of a full-fletched pure and reliable EFT calculation one might wonder if it is possible to 
obtain a model independent result by combining the low-energy chiral constraints with experimentally obtained reaction data and some general principles of quantum 
field theory and/or QCD. Indeed dispersion theory offers such a framework and is of practical use in energy regimes where 
the number of open channels is not too large \cite{Donoghue:1990xh,Ananthanarayan:2000ht,GarciaMartin:2011cn}. 
However, for the $\eta'$-$\pi$ to $\eta$-$\pi$ scattering amplitude 
a recent work \cite{schneiderPhD} in that direction has revealed large uncertainties based on the poor data situation 
concerning the $\eta$-$\pi$ channel. The work of \cite{schneiderPhD} has focussed on the decay kinematics 
$\eta' \to \eta \pi \pi$ where at least the kaon channel is still closed. For our situation at hand, $\eta'$-$\pi$ scattering, 
the situation is even worse and one has to face a formidable coupled-channel problem involving at least the channels 
$\eta'$-$\pi$, $\eta$-$\pi$, $K$-$\bar K$, if not $3\pi$. The better known part of this problem is the t-channel, i.e.\ formally 
the reaction $\eta' \eta$ to $2\pi$, where detailed information is available for the 
pion phase shifts \cite{Ananthanarayan:2000ht,GarciaMartin:2011cn}. 
We have described this part just by the exchange of one $f_0$ while in the real world of three colors one finds a broad resonance 
at around 500 MeV and a second one at the kaon threshold \cite{Ananthanarayan:2000ht,GarciaMartin:2011cn,Agashe:2014kda}. 
Thus it might be interesting for future investigations to replace our tree-level large-$N_c$ calculation by an approach 
that combines measured pion phase shifts (t-channel) with a resonance modelling in the 
$a_0$ channel (s- and u-channel). Still we 
regard the chiral constraints obtained from large-$N_c$ $ChPT$ as a vital input to such an approach. 

In the present work we have focused on {\em thermal} changes of a property of the $\eta'$ meson. Of course, it is also interesting to look at {\em nuclear} modifications. This has been addressed, for instance, in \cite{Oset:2010ub,Jido:2011pq,Itahashi:2012ut,Nanova:2012vw,Sakai:2013nba,Nanova:2013fxl}. Our analysis based on mesonic $ChPT$  adds the following new aspect to the corresponding nuclear calculations: Significant in-medium effects in a nuclear environment can be revealed by starting with thermal, i.e.\ pion induced effects and coupling nucleonic and $\Delta$ degrees of freedom to the pion legs \cite{Herrmann:1993za}. Starting with the processes studied in the present work, this suggests that the following elementary reactions are important for the nuclear modifications of the mass and width of an $\eta'$ meson: $\eta' N \to \eta \pi N, \eta \pi \Delta, K \bar K N,  K \bar K \Delta$. Obviously these reactions concern final states with three particles. (If one considers the final decay of the $\Delta$ baryon, one involves even four-body states.) In contrast many works in the literature rely on the dominance of two-body reactions. However, with a threshold of nearly 2 GeV the reaction of an $\eta'$ meson on a nucleon provides ample of phase space for many-body final states. It remains to be seen if there is a significant or even dominant impact of the three- and four-body final states on the nuclear modifications of the properties of the $\eta'$ meson.

\section*{Acknowledgments}
S.L.\ thanks V.\ Metag for stimulating his interest in the $\eta'$ meson. He also acknowledges valuable discussions with 
D.\ Jido, S.H.\ Lee, M.\ Nanova, E.\ Oset, and C.\ Wilkin.

\appendix 
\section{Matrix elements $\mathcal{M}_{\eta'\pi\rightarrow\eta\pi}$ and $\mathcal{M}_{\eta'\pi\rightarrow K \bar{K}}$ in \textit{ChPT}} \label{A}
Starting from the effective Lagrangian~\eqref{effL} we obtain the matrix elements, which are presented here in terms of the physical fields:
\begin{equation}\label{etapi}\begin{split}
\mathcal{M}_{\eta'\pi\rightarrow \eta\pi}^{ChPT}(s,t,u) &= \frac{c_1}{F^2} \bigg[\frac{M_{\pi}^2}{2}\\&\hspace{3em}+\frac{2(3L_2+L_3)}{F_{\pi}^2}\left(s^2+t^2+u^2-M_{\eta'}^{4}-M_{\eta}^{4}-2M_{\pi}^{4}\right)\\& \hspace{3em} -\frac{2L_5}{F_{\pi}^2}\left(M_{\eta'}^2+M_{\eta}^2+2M_{\pi}^2\right)M_{\pi}^2 +\frac{24L_8}{F_{\pi}^2}M_{\pi}^4+\frac{2}{3}\Lambda_{2} M_{\pi}^2\bigg] \\&\phantom{=}+\frac{c_2}{F^2} \frac{\sqrt{2}}{3}\Lambda_{2} M_{\pi}^2,
\end{split}\end{equation}
where in accordance with \citep{Escribano:2010wt} we have defined:
\begin{equation}\label{c1c2}\begin{aligned}
&c_1 =-\frac{F^2}{3F_8^2F_0^2\cos ^2(\theta_8 -\theta_0)}\Big[2F^2_8\sin(2\theta_8)-F^2_0\sin(2\theta_0)\\ &\hspace{12em} -2\sqrt{2}F_8F_0\cos(\theta_8 +\theta_0)\Big],\\
&c_2 =-\frac{F^2}{3F_8^2F_0^2\cos ^2(\theta_8 -\theta_0)}\Big[\sqrt{2}F^2_8\sin(2\theta_8)+\sqrt{2}F^2_0\sin(2\theta_0)\\& \hspace{12em} +F_8F_0\cos(\theta_8 +\theta_0)\Big].
\end{aligned}\end{equation}
Expression~\eqref{etapi} is the NLO result. The LO expression is:
\begin{equation}\label{LOetapi}\begin{split}
(\mathcal{M}_{\eta'\pi\rightarrow \eta\pi})_{LO} &= \frac{c_1M^2_\pi}{2F^2} \\& =\frac{M^2_\pi}{6F^2}\left[2\sqrt{2}\cos(2\theta)-\sin(2\theta)\right],
\end{split}\end{equation}
since at LO $F_0=F_8=F$ and $\theta_0=\theta_8=\theta$, i.e there is only one mixing angle which has the approximate value of $-20^{\circ}$.

The kaonic matrix element is:
\begin{equation}\label{kaons}\begin{split}
\mathcal{M}_{\eta'\pi\rightarrow K\bar{K}}(s,t,u) =&\, \frac{c_3F_8\cos\theta_{8}}{3F_K^2F_{\pi}F^2} \bigg\{F^2(2M_K^2+M_{\pi}^2)(1+\Lambda_2)\\& \hspace{2em} +12(3L_2+L_3)(s^2+t^2+u^2-M_{\eta'}^{4}-M_{\pi}^{4}-2M_{K}^{4})\\&\hspace{2em} -4L_5\big[2M_{\pi}^2(M_{\eta'}^2+3M_K^2)+M_K^2(M_{\eta'}^2+3M_{\pi}^2)\\&\hspace{5em} +3s(M_K^2-M_{\pi}^2)-2(2M_K^4+M_{\pi}^4)\big]\\&\hspace{2em} +16L_8(2M_K^4-M_{\pi}^4+8M_K^2M_{\pi}^2)\bigg\}\\&+  \frac{c_3\sqrt{2}F_0\sin\theta_{0}}{3F_K^2F_{\pi}F^2} \bigg\{-\frac{F^2}{4}(3M_{\eta'}^2+8M_K^2+M_{\pi}^2-9s) \\& \hspace{2em}+3L_3\big[2tu-s(t+u)-2M_K^4\\&\hspace{5em}  -2M_{\pi}^2(M_{\eta'}^2-M_K^2)+2M_K^2M_{\eta'}^2\big]\\&\hspace{2em} -2L_5\big[7M_{\eta'}^2M_K^2-M_{\eta'}^2M_{\pi}^2+9M_K^2M_{\pi}^2\\&\hspace{5em} -3s(5M_K^2+M_{\pi}^2)+8M_K^4+M_{\pi}^4\big]\\& \hspace{2em}-16L_8(2M_K^4-M_{\pi}^4-M_K^2M_{\pi}^2) \bigg\},
\end{split}\end{equation}
where
\begin{equation}\label{c3}
c_3=\frac{F^2}{\sqrt{3}F_0F_8\cos(\theta_8 -\theta_0)}.
\end{equation}
The LO expression is:
\begin{equation}\label{LOkaons}\begin{split}
(\mathcal{M}_{\eta'\pi\rightarrow K\bar{K}})_{LO} &= \frac{1}{6\sqrt{6}F^2}\Big[2\sqrt{2}(2M^2_K+M^2_\pi)\cos\theta \\& \hspace{5em}-(3M^2_{\eta'}+8M^2_K+M^2_\pi-9s)\sin\theta\Big].
\end{split}\end{equation}
\subsection*{Scale independence of matrix elements}
The matrix elements~\eqref{etapi} and ~\eqref{kaons} contain among the LEC:s also $\Lambda_2$ and $F_0$ which have a renormalization scale dependence through the parameter $\Lambda_3$. Since we want to calculate physical quantities such as cross sections, we want to be sure that in the matrix elements only scale invariant combinations of these parameters appear. 

Beginning with $\mathcal{M}_{\eta'\pi\rightarrow\eta\pi}$ we note that every term is multiplied by either $c_1$ or $c_2$ and therefore contains $F_0$. According to~\eqref{values}, we can write $F_0=(1+\Lambda_3)\tilde{F_0}$ where $\tilde{F_0}$ is a scale independent number and replace $F_0$ with $\tilde{F_0}$ in the NLO parts. In fact, up to corrections of $O(\delta^2)$  we only need to consider the $\Lambda_3$ dependence of the very first contribution in~\eqref{etapi} which comes from the LO Lagrangian. This is given by:
\begin{equation*}\begin{split}
(\mathcal{M}_{\eta'\pi\rightarrow \eta\pi})_{LO}=
-\frac{M^2_\pi}{2}\frac{1}{3\cos^2(\theta_8-\theta_0)}\bigg(\frac{2}{F_0^2}\sin(2\theta_8)&-\frac{1}{F_8^2}\sin(2\theta_0)\\&-\frac{2\sqrt{2}}{F_8F_0}\cos(\theta_8+\theta_0)\bigg).
\end{split}\end{equation*}
Expanding $1/F_0\approx (1-\Lambda_3)/\tilde{F_0}$ and  $1/F_0^2\approx (1-2\Lambda_3)/\tilde{F_0}^2$ gives
\begin{equation*}\begin{split}
-\frac{M^2_\pi}{2}\frac{1}{3\cos^2(\theta_8-\theta_0)}\bigg(\frac{2(1-2\Lambda_3)}{\tilde{F_0}^2}\sin(2\theta_8)&-\frac{1}{F_8^2}\sin(2\theta_0)\\&-\frac{2\sqrt{2}(1-\Lambda_3)}{F_8\tilde{F_0}}\cos(\theta_8+\theta_0)\bigg),
\end{split}\end{equation*}
from which we can read off the part proportional to $\Lambda_3$:
\begin{equation}\label{lambda3}
-\Lambda_3\frac{\sqrt{2}M^2_\pi}{3\cos^2(\theta_8-\theta_0)}\left(-\frac{\sqrt{2}}{\tilde{F_0}^2}\sin(2\theta_8)+\frac{1}{F_8\tilde{F_0}}\cos(\theta_8+\theta_0)\right).
\end{equation}
The other scale dependent part of $\mathcal{M}_{\eta'\pi\rightarrow\eta\pi}$ is the one proportional to $\Lambda_2$:
\begin{equation*}
\Lambda_2\frac{\sqrt{2}M^2_\pi}{3F^2}\left(\sqrt{2}c_1+c_2\right)\,.
\end{equation*}
Writing explicitly the expressions for the coefficients $c_1$ and $c_2$~\eqref{c1c2} and using the approximation $F_0\approx \tilde{F_0}$ we obtain:
\begin{equation}
-\Lambda_2\frac{\sqrt{2}M^2_\pi}{3\cos^2(\theta_8-\theta_0)}\left(\frac{\sqrt{2}}{\tilde{F_0}^2}\sin(2\theta_8)-\frac{1}{F_8\tilde{F_0}}\cos(\theta_8+\theta_0)\right).
\end{equation}
The above is exactly the negative of~\eqref{lambda3}. This means that up to corrections of $O(\delta^2)$ the scale dependent terms only show up in the scale invariant combination $\Lambda_2-\Lambda_3$.

Regarding the kaonic matrix element~\eqref{kaons}, note that the part proportional to $\sin\theta_0$ is already scale independent. The part proportional to $\cos\theta_8$ contains a term proportional to $\Lambda_2$:
\begin{equation*}
\frac{c_3F_8\cos\theta_8}{3F^2_KF_{\pi}F^2}\Lambda_2F^2\left(2M^2_K+M^2_\pi\right)\,.
\end{equation*}
Regarding the other NLO terms we can just replace $F_0$ with $\tilde{F_0}$. The last thing to analyze is the LO part:
\begin{equation*}
\frac{c_3F_8\cos\theta_8}{3F^2_KF_{\pi}F^2}F^2\left(2M^2_K+M^2_\pi\right)\,.
\end{equation*}
Putting together these two terms we obtain:
\begin{equation*}\label{LOelambda}
\frac{c_3F_8\cos\theta_8}{3F^2_KF_{\pi}F^2}F^2\left(2M^2_K+M^2_\pi\right)(1+\Lambda_2)\,.
\end{equation*}
From the definition of $c_3$~\eqref{c3} and by expanding $1/F_0$, the above expression becomes:
\begin{equation*}
\frac{\cos\theta_8}{3\sqrt{3}F^2_KF_\pi\cos(\theta_8-\theta_0)}\frac{1}{\tilde{F_0}}F^2\left(2M^2_K+M^2_\pi\right)(1-\Lambda_3)(1+\Lambda_2)\,.
\end{equation*}
Since we have that
\begin{equation*}
(1-\Lambda_3)(1+\Lambda_2)=1+\Lambda_2-\Lambda_3+O(\delta^2),
\end{equation*}
we see that also in the kaonic matrix element only the scale invariant combination $\Lambda_2-\Lambda_3$ appears.

\pagebreak

\section{Matrix elements $\mathcal{M}_{\eta'\pi\rightarrow\eta\pi}$ and $\mathcal{M}_{\eta'\pi\rightarrow K \bar{K}}$ in \textit{RChT}} \label{B}
From the $RChT$ Lagrangian~\eqref{RChTlag} we can find the two matrix elements of interest. We will focus first on the matrix element associated to the reaction $\eta'\pi\rightarrow\eta\pi$. In the notation of~\citep{Escribano:2010wt} we have:
\begin{equation}\label{RChTetapi}\begin{split}
\mathcal{M}_{\eta'\pi\rightarrow \eta\pi}^{RChT}&=\frac{c_{1}}{F^2}\biggl[\frac{M^2_\pi}{2}\\&+\frac{1}{F^2_\pi}\frac{\big(c_d(s-M^2_{\eta'}-M^2_\pi)+2c_m M^2_\pi\big)\big(c_d(s-M^2_\eta-M^2_\pi)+2c_mM^2_\pi\big)}{M^2_S-s}\\&+\frac{1}{F^2_\pi}\frac{\big(c_d(t-M^2_{\eta'}-M^2_\eta)+2c_mM^2_\pi\big)\big(c_d(t-2M^2_\pi)+2c_mM^2_\pi\big)}{M^2_S-t}\\&+\frac{1}{F^2_\pi}\frac{\big(c_d(u-M^2_{\eta'}-M^2_\pi)+2c_m M^2_\pi\big)\big(c_d(u-M^2_\eta-M^2_\pi)+2c_m M^2_\pi\big)}{M^2_S-u}\\&+ \frac{2}{3}\Lambda_2M_\pi^2\biggr]+\frac{c_2}{F^2}\frac{\sqrt{2}}{3}\Lambda_2M_\pi^2,
\end{split}\end{equation}
where the isovector scalar $a_0$ is exchanged in the s- and u-channel and the scalar $f_0$ in the t-channel. We recognize the LO contribution (the very first term in~\eqref{RChTetapi}) and the $\Lambda_2$ terms from $ChPT$ (see~\eqref{etapi}). The part with the resonance propagators in~\eqref{RChTetapi} replaces the LEC:s parts $\sim L_i$ of the $ChPT$ expression. 

Another interesting aspect is that in the scattering amplitude~\eqref{RChTetapi} the point-interaction diagram $\sim c_m^2$ coming from~\eqref{Lstilde} does not show up. To understand why, we have to go back to the LO scattering amplitude~\eqref{LOetapi} which is:
\begin{equation}\label{extraterm}
(\mathcal{M}_{\eta'\pi\rightarrow \eta\pi})_{LO} = \frac{c_1 M_{\pi LO}^2 }{2F^2},
\end{equation}
where at LO we were allowed to make the approximation: 
\begin{equation*}
\frac{2BmF^2}{F^2_\pi}=M_{\pi LO}^2\,.
\end{equation*}
However when we consider the scattering amplitude up to (including) NLO, the above approximation is not accurate enough. We have to use instead the relations~\eqref{system}. In particular from
\begin{equation*}
F^2_\pi M^2_\pi =2F^2Bm+64(Bm)^2L_8
\end{equation*}
and recalling~\eqref{LECrel}
\begin{equation*}
L_8=\frac{c_m^2}{2M^2_S},
\end{equation*}
we can make the following replacement in~\eqref{extraterm}:
\begin{equation}
M_{\pi LO}^2=M_{\pi}^2 -\frac{8c_m^2M_\pi^4}{F_\pi^2M^2_S}\,.
\end{equation}
We thus get an extra term $\sim c_m^2$ that must be included at NLO. This additional term cancels the contact term coming from~\eqref{Lstilde}.

In the heavy scalar mass limit, that is $s,t,u \ll M_S^2$, the amplitude~\eqref{RChTetapi} becomes:
\begin{equation}\begin{split}
\mathcal{M}_{\eta'\pi\rightarrow \eta\pi}^{RChT}\rightarrow &\,\frac{c_{1}}{F^2}\biggl[\frac{M^2_\pi}{2}+\frac{c_d^2}{F^2_\pi M^2_S}(s^2+t^2+u^2-M_{\eta'}^4-M_{\eta}^4-2M_{\pi}^4)\\ &\qquad -\frac{2c_dc_m}{F^2_\pi M^2_S}(M_{\eta'}^2+M_{\eta}^2+2M_{\pi}^2)M^2_\pi+\frac{12c_m^2}{F^2_\pi M^2_S}M_\pi^4+ \frac{2}{3}\Lambda_2M_\pi^2\biggr]\\&+\frac{c_2}{F^2}\frac{\sqrt{2}}{3}\Lambda_2M_\pi^2,
\end{split}\end{equation}
which turns out to resemble the large-$N_c\  ChPT$ amplitude~\eqref{etapi}. After using the LEC:s relations~\eqref{LECrel}, the two amplitudes are seen to be identical. This shows that in the heavy mass limit $RChT$ reduces to $ChPT$, which means that $ChPT$ is fully recovered at low energies. 

The matrix element with two kaons in the final state is: 
\begin{equation}\label{kaonRChT}\begin{split}
\mathcal{M}_{\eta'\pi\rightarrow K\bar{K}}^{RChT}(s,t,u) &= \frac{c_3}{3F_K^2F_{\pi}F^2} \biggl[F^2F_8\cos\theta_{8}(2M_K^2+M_{\pi}^2)(1+\Lambda_2)\\& \hspace{6em}-\frac{\sqrt{2}F^2F_0\sin\theta_{0}}{4}(3M_{\eta'}^2+8M_K^2+M_{\pi}^2-9s)\\&\hspace{6em}+\frac{f_1(s)}{s-M^2_S}+\frac{f_2(t)}{t-M^2_S}+\frac{f_2(u)}{u-M^2_S}\\&\hspace{6em}-\frac{f_3(s,t,u)}{t-M^2_V}-\frac{f_3(s,u,t)}{u-M^2_V}-\frac{g(s)}{M_S^2}\biggr],
\end{split}\end{equation}
with
\begin{equation*}\begin{split}
f_1(x)=3 \left[2 M_K^2 (c_d-c_m)-c_d x\right] &\left[2 c_m M_\pi^2-c_d \left(M_{\eta'}^2+M_\pi^2-x\right)\right]\\& \times
   \left(\sqrt{2} F_0 \sin \theta_0+2 F_8 \cos \theta_8\right),
\end{split}\end{equation*}
\begin{equation*}\begin{split}
f_2(x)=&\,\frac{3}{2}\big[(c_d-c_m) (M_K^2+M_\pi^2)-c_d x\big] \Big[\sqrt{2} F_0 \sin \theta_0\big(c_d
   (M_{\eta'}^2+M_K^2-x)\\&-5 c_m M_K^2+3 c_m M_\pi^2\big)-4 F_8 \cos \theta_8 \big(c_d
   (M_{\eta'}^2+M_K^2-x)-2 c_m M_K^2\big)\Big],
\end{split}\end{equation*}
\begin{equation*}\begin{split}
f_3(s,x,y)=\frac{9 F_0 G_V^2}{2
   \sqrt{2}} \sin \theta_0 \left[(M_K^2-M^2_{\eta'}) (M_K^2-M^2_\pi) +x (s-y)\right]
\end{split}\end{equation*}
and
\begin{equation*}\begin{split}
g(x)=\,c_m &\Big[\sqrt{2} F_0 \sin \theta_0 \Big(c_d \big(16
   M_K^4+M_K^2 (2 M_{\eta'}^2+6 M_\pi^2-15 x) \\&+M_\pi^2 (4 M_{\eta'}^2-4 M_\pi^2-3 x)\big)+c_m (M_K^2-M_\pi^2)^2\Big)\\&-2 F_8 \cos
   \theta_8 \Big(c_d \big(8 M_K^4+M_K^2 (M_{\eta'}^2+3 M_\pi^2-3 x)\\&+M_\pi^2 (-M_{\eta'}^2+ M_\pi^2+3 x)\big)-4
   c_m (M_K^2-M_\pi^2)^2\Big)\Big].
\end{split}\end{equation*}

In the expression for the kaonic matrix element we recognize the LO part and the $\Lambda_2$ term. All the rest is given in terms of functions of the Mandelstam variables $s,\ t$ and $u$. In particular the $f_i$ come from the tree-level exchanges of the scalar and vector resonances. The corresponding denominators indicate in which channel the resonance is exchanged together with its nature (vector $M_V$, scalar $M_S$). The function $g$ comes from point-interaction diagrams, due to the two extra terms obtained as a consequence of the shift in the $S$ field (see~\eqref{Lstilde}). Note that the scattering amplitude contains also the additional NLO terms obtained from the LO expression using the relations~\eqref{system}.    

Exactly as it was for the $\eta\pi$ final state case, the matrix element~\eqref{kaonRChT} reduces to the correspondent $ChPT$ amplitude~\eqref{kaons} in the low-energy limit. One can formally see this by taking the limit $M^2_R\gg s,t,u$ (so that the energy dependence in the denominators disappears) and using the LEC:s relations~\eqref{LECrel}.



\bibliographystyle{model1-num-names}
\bibliography{Bibliography}







\end{document}